\documentclass[a4paper,11pt]{article}
\usepackage[toc,page]{appendix}
\usepackage{jheppub}
\usepackage{setspace}
\usepackage{amssymb}
\usepackage{amsmath,amsfonts}
\usepackage{graphicx}
\usepackage{mathrsfs}
\usepackage[vcentermath]{youngtab}
\usepackage{color}
\numberwithin{equation}{section}
\usepackage{bbm}

\newcommand{\be}{\begin{equation}}
\newcommand{\ee}{\end{equation}}
\newcommand{\bea}{\begin{eqnarray}}
\newcommand{\eea}{\end{eqnarray}}

\newcommand{\ket}[1]{\left\lvert #1 \right\rangle}

\usepackage{multirow}

\newcommand{\commute}[2]{\left[ #1 \, , \, #2 \right]}
\newcommand{\anticommute}[2]{\left\{ #1 \, , \, #2 \right\}}
\usepackage{array}
\usepackage{graphicx}
\usepackage{longtable}
\usepackage{color}
\usepackage{bbm}
\usepackage{appendix}

\newcommand{\bep}{\begin{picture}}
\newcommand{\eep}{\end{picture}}

\newcounter{YoungHeight}\newcounter{YoungWidth}

\newcounter{Mul1}\newcounter{Mul2}\newcounter{Mul3}\newcounter{Mul4}
\newcounter{A1}\newcounter{A2}
\newcounter{B3}
\newcounter{C3}\newcounter{C4}

\newcounter{T0}\newcounter{T1}

\newcounter{R0}

\newlength{\txtHShift}

\newlength{\txtWidth}

\newcommand{\Add}[3]{\setcounter{#1}{#2}\addtocounter{#1}{#3}}

\newcommand{\Length}[1]{#10}
\newcommand{\YoungScale}{}

\newcommand{\BlockA}[2]{{\YoungScale\bep(\Length{#1},\Length{#2}){\Add{A1}{#1}{1}\Add{A2}{#2}{1}}%
\multiput(0,0)(10,0){\value{A1}}{\line(0,1){\Length{#2}}}\multiput(0,0)(0,10){\value{A2}}{\line(1,0){\Length{#1}}}%
\setcounter{YoungHeight}{\Length{#2}}\setcounter{YoungWidth}{\Length{#1}}\eep}}

\newcommand{\YoungB}{\BlockA{2}{1}}

\newcommand{\YoungAA}{\BlockA{1}{2}}

\newcommand{\YoungBB}{\BlockA{2}{2}}

\newcommand{\YoungAAAA}{\BlockA{1}{4}}

\usepackage[normalem]{ulem}

\begin{document}

\title{Minimal unitary representation of $5d$ superconformal algebra $F(4)$ and $AdS_6/CFT_5$ higher spin (super)-algebras}

\author{Sudarshan Fernando$^a$  and  } 
\author{Murat G\"unaydin$^{b}$}
 
\affiliation{$^{a}$  Physical Sciences Department \\
Kutztown University \\
Kutztown, PA 19530, USA}
\affiliation{$^b$ Institute for Gravitation and the Cosmos \\
 Physics Department ,
Pennsylvania State University\\
University Park, PA 16802, USA}

\emailAdd{fernando@kutztown.edu}
\emailAdd{murat@phys.psu.edu}
\abstract{
We study the minimal unitary representation (minrep) of $SO(5,2)$, obtained by quantization of its geometric quasiconformal action, its deformations and supersymmetric extensions.  The minrep of $SO(5,2)$ describes a massless conformal scalar field in five dimensions and admits a unique ``deformation'' which describes a massless conformal spinor. Scalar and spinor minreps of $SO(5,2)$ are the $5d$ analogs of Dirac's singletons of $SO(3,2)$. We then construct the minimal unitary representation of the unique $5d$ superconformal algebra $F(4)$ with the even subalgebra $SO(5,2) \times SU(2)$. The minrep of $F(4)$ describes a massless conformal supermultiplet consisting of two scalar and one spinor fields. We then extend our results to the construction of higher spin $AdS_6/CFT_5$  (super)-algebras. The Joseph ideal of the minrep of $SO(5,2)$ vanishes identically as  operators and hence its enveloping algebra yields the $AdS_6/CFT_5$ bosonic higher spin algebra directly. The enveloping algebra of the spinor minrep  defines a ``deformed'' higher spin algebra for which a deformed Joseph ideal vanishes identically as  operators. These results are then extended to the construction of the unique higher spin $AdS_6/CFT_5$ superalgebra as the enveloping algebra of the minimal unitary realization of $F(4)$ obtained by the quasiconformal methods.}

\maketitle

\section{Introduction}

In previous work we gave a construction of the minimal unitary realization of the Lie algebra of four-dimensional conformal group $SU(2,2)$ and showed that it admits a one-parameter family of deformations \cite{Fernando:2009fq}. The minimal unitary representation (minrep) and its deformations describe all the massless conformal fields in four dimensions, and the deformation parameter is simply the helicity which can take on continuous values. These results were also extended to the superalgebras $SU(2,2|N)$, and it was shown that minimal unitary supermultiplet and its deformations describe massless $N$-extended superconformal multiplets in four dimensions. For $PSU(2,2|4)$, the minimal unitary supermultiplet is simply the $N=4$ Yang-Mills supermultiplet \cite{Fernando:2009fq}.

Minimal unitary representation of six-dimensional conformal group $SO(6,2) \approx SO^*(8)$ and its deformations and their supersymmetric extensions were studied in \cite{Fernando:2010dp,Fernando:2010ia}. Particularly in \cite{Fernando:2010dp}, it was shown that the minimal unitary representation of $SO(6,2)$ corresponds to a massless conformal scalar field in six dimensions, and it admits a discrete infinite family of deformations, labelled by the spin $t$ of an $SU(2)$ subgroup of the little group $SO(4)$ of massless particles, which is the six-dimensional analog of four-dimensional helicity. This infinite family of unitary irreducible representations of $SO^*(8)$ are isomorphic to the doubleton irreps of $SO^*(8)$ constructed in \cite{Gunaydin:1984wc,Gunaydin:1999ci} using covariant twistorial oscillators. These results were also extended to six-dimensional conformal supergroups $OSp(8^*|2N)$ with the even subgroup $SO^*(8) \times U\!Sp(2N)$ in \cite{Fernando:2010ia}. The minimal unitary supermultiplet of $OSp(8^*|2N)$ also admits a discrete infinite family of deformations, labelled by the spin $t$ of the $SU(2)$ subgroup that describe massless conformal supermultiplets in six dimensions. For $OSp(8^*|4)$, the minimal unitary supermultiplet is the $(2,0)$ conformal supermultiplet that was first constructed in \cite{Gunaydin:1984wc}. The existence of an interacting non-gravitational theory of $(2,0)$ conformal supermultiplets that arise in a singular limit of ten-dimensional type IIB string theory with a codimension four singularity of ADE-type was predicted by Witten \cite{Witten:1995zh}. 

The Kaluza-Klein spectrum of IIB supergravity over $AdS_5 \times S^5$ was first obtained back in 1984 in \cite{Gunaydin:1984fk} by tensoring the CPT self-conjugate doubleton supermultiplet of $PSU(2,2|4)$ repeatedly with itself and restricting to the CPT self-conjugate short supermultiplets. The authors of \cite{Gunaydin:1984fk} also pointed out that the CPT self-conjugate doubleton of $PSU(2,2|4)$ does not admit a Poincar\'{e} limit in $AdS_5$, and its field theory that lives on the boundary is the four-dimensional $N=4$ super Yang-Mills theory which was known to be conformally invariant. Later Maldacena proposed the duality between type IIB superstring theory on $AdS_5 \times S^5$ and the $N=4$ $SU(\mathcal{N})$ super Yang-Mills theory in the large $\mathcal{N}$ limit \cite{Maldacena:1997re}.

Again back in 1984, the spectrum of eleven-dimensional supergravity was obtained by tensoring the $(2,0)$ supermultiplet of $OSp(8^*|4)$, and it was pointed out that the field theory of the scalar doubleton supermultiplet lives on the boundary of $AdS_7$ as a conformally invariant field theory \cite{Gunaydin:1984wc}. The interacting six-dimensional conformal theory of $(2,0)$ multiplets  is also believed to be dual to M-theory over $AdS_7 \times S^4$ \cite{Maldacena:1997re}. 

The fact that the $N=4$ Yang-Mills supermultiplet and the $(2,0)$ supermultiplet are the minimal unitary supermultiplets of $PSU(2,2|4)$ and $OSp(8^*|4)$, respectively, shows their fundamental importance from a mathematical point of view as well.

More recently the results of \cite{Fernando:2009fq,Fernando:2010dp,Fernando:2010ia} were reformulated in terms of deformed twistors in four and six dimensions that transform nonlinearly under the Lorentz group and applied to higher spin theories in $AdS_5$ and $AdS_7$ \cite{Govil:2013uta,Govil:2014uwa}.  One finds that the enveloping algebras of the minimal unitary realizations of $SU(2,2)$ and $SO^*(8)$, obtained by the quasiconformal methods, yield directly the higher spin algebras in $AdS_5$ and $AdS_7$. The bosonic higher spin algebras in $AdS_d$, as studied by Fradkin, Vasiliev and their collaborators \cite{Fradkin:1986qy,Konshtein:1988yg}, correspond to the enveloping algebras of the $AdS_d$ Lie algebras $SO(d-1,2)$ quotiented by a two sided ideal. That the Fradkin-Vasiliev higher spin algebra in $AdS_4$ is simply the enveloping algebra of the singletonic realization of $SO(3,2)$ was pointed out long time ago in \cite{Gunaydin:1989um}. Vasiliev showed that the enveloping algebra of $AdS_d$ group $SO(d-1,2)$ must be quotiented by the ideal that is the annihilator of the scalar singleton representation \cite{Vasiliev:1999ba}. Later Eastwood identified this ideal to be the Joseph ideal of the minimal unitary representation of $SO(d-1,2)$ \cite{Eastwood:2002su}.

The quasiconformal realization of the minimal unitary representation of  symplectic groups coincides with the realization as bilinear of oscillators \cite{Gunaydin:2006vz} and the Joseph ideal vanishes identically \cite{Govil:2013uta}. That is why the enveloping algebra of the singletonic realization of $Sp(4,\mathbb{R}) \approx SO(3,2)$ as bilinears of covariant twistorial oscillators leads directly to the Fradkin-Vasiliev higher spin algebra in $AdS_4$. For the doubletonic realization of the minimal unitary representation of $AdS_5$ group $SO(4,2)$ and $AdS_7$ group $SO(6,2)$, in terms of bilinears of covariant twistorial oscillators, the Joseph ideal does not vanish identically as an operator. However for their minreps obtained by quasiconformal methods, the Joseph ideal vanishes identically as an operator \cite{Govil:2013uta,Govil:2014uwa} both in $AdS_5$ and $AdS_7$, respectively. Hence their enveloping algebras yield directly the corresponding bosonic higher spin algebras. Similarly for each deformed minrep there is a deformed ideal that vanishes identically. Taking their enveloping algebras lead to a continuous one-parameter family of higher spin algebras and superalgebras in $AdS_5$ and a discrete infinite family of higher spin algebras and superalgebras in $AdS_7$ \cite{Govil:2013uta,Govil:2014uwa}.

In this paper we extend the above mentioned results to $AdS_6$/$CFT_5$ group $SO(5,2)$ and its \emph{unique} supersymmetric extension, namely the exceptional supergroup $F(4)$ with the even subgroup $SO(5,2) \times SU(2)$. We shall first review the construction of the minrep of $SO(5,2)$ by quantization of its geometric realization as a quasiconformal group following \cite{Gunaydin:2006vz}. We show that it corresponds to a conformal massless scalar field in five dimensions. We then study the possible deformations of the minrep and find only \emph{a single deformation} corresponding to a conformally massless spinor field in five dimensions. This is similar to the situation in three dimensions, where the only conformally massless fields are a scalar and a spinor field corresponding to Dirac's singletons. We then study the minimal unitary realization of the \emph{unique} simple conformal supergroup $F(4)$ with the even subgroup $SO(5,2) \times SU(2)$. We find that the spinor minrep, together with two copies of the scalar minrep of $SO(5,2)$, form the minimal unitary supermultiplet of $F(4)$. We then extend these results to the construction of $AdS_6/CFT_5$ algebras and superalgebra. We show that the Joseph ideal vanishes identically as  operators for the minrep  of $SO(5,2)$  and its enveloping algebra yields the bosonic higher spin algebra of Vasiliev-type in $AdS_6$. For the deformed minrep a certain deformation of the Joseph ideal vanishes and its enveloping algebra yields a deformed higher spin algebra. The enveloping algebra of the superalgebra $F(4)$ yields the \emph{unique} higher spin superalgebra in $AdS_6$\footnote{By an abuse of notation we shall  use $F(4)$ to denote both the supergroup as well as  its superalgebra $\mathfrak{f}(4)$ with the even subalgebra $\mathfrak{so}(5,2)\oplus \mathfrak{su}(2)$.} .

The plan of the paper is as follows. In section \ref{sec:geomSO(5,2)}, we construct the geometric realization of $SO(5,2)$ as a quasiconformal group, following the method outlined in \cite{Gunaydin:2005zz}.
Then in section \ref{sec:minrepSO(5,2)}, we obtain the minimal unitary representation of $SO(5,2)$ via the quantization of the geometric quasiconformal action and show that there is a two-parameter family of degree two polynomials of $\mathfrak{so}(5,2)$ generators that reduces to a $c$-number according to Joseph's theorem \cite{Joseph:1974}.
We present the 3-grading of $\mathfrak{so}(5,2)$ with respect to the noncompact subalgebra $\mathfrak{so}(4,1) \oplus \mathfrak{so}(1,1)$ in section \ref{sec:SO(5,2)nc3G}, and with respect to the compact subalgebra $\mathfrak{so}(5) \oplus \mathfrak{so}(2) \approx \mathfrak{ups}(4) \oplus \mathfrak{u}(1)$ in section \ref{sec:SO(5,2)c3G}. In that section, the results are presented separately in $SO(5)$-covariant and $U\!Sp(4)$-covariant forms.
Then in section \ref{sec:SU(1,1)_K} we discuss the properties of a distinguished $SU(1,1)$ subgroup of $SO(5,2)$ generated by singular (isotonic) oscillators.
We then give the K-type decomposition of the minrep of $SO(5,2)$ in section \ref{sec:undeformedminrep} and show that it corresponds to a massless conformal scalar field in five dimensions.
We study  the deformations of the minrep in section \ref{sec:deformedSO(5,2)} and show that it admits a unique deformation and  give its K-type decomposition in section \ref{sec:deformedminrep}.
In section \ref{sec:F(4)minrep}, we construct the minimal unitary representation of the unique exceptional  superconformal algebra $\mathfrak{f}(4)$ in five dimensions.
A 3-grading of $\mathfrak{f}(4)$ with respect to the compact subsuperalgebra $\mathfrak{osp}(2|4) \oplus \mathfrak{u}(1)$ is given in section \ref{sec:F(4)c3G}, and the supermultiplet of conformal fields corresponding to the minrep  of $\mathfrak{f}(4)$ is given in section \ref{sec:F(4)supermultiplet}.
Then in section \ref{sec:bosonicHS}, we show that the Joseph ideal vanishes identically as an operator for the minrep of $SO(5,2)$ and  hence  its  universal enveloping algebra yields directly the bosonic $AdS_6/CFT_5$ higher spin algebra. Similarly for the deformed minrep of $SO(5,2)$ a certain deformation of the Joseph ideal vanishes identically and its enveloping algebra yields a deformed $AdS_6/CFT_5$ higher spin algebra. 
Finally in section \ref{sec:higherspin} we define the unique  $AdS_6/CFT_5$  higher spin superalgebra as the enveloping algebra of the minrep of  $F(4)$ and study some of its properties followed by some concluding comments.


\section{Geometric realization of $SO(5,2)$ as a quasiconformal group}
\label{sec:geomSO(5,2)}

In this section we shall review the geometric quasiconformal realization of $SO(5,2)$ following \cite{Gunaydin:2005zz}. The Lie algebra $\mathfrak{so}(5,2)$ admits the following 5-grading with respect to its subalgebra $\mathfrak{so}(1,1) \oplus \mathfrak{so}(3) \oplus \mathfrak{sp}(2,\mathbb{R})$: 
\begin{equation}
\mathfrak{so}(5,2)
= \mathbf{1}^{(-2)} \oplus
  \left( \mathbf{3} , \mathbf{2} \right)^{(-1)} \oplus
  \left[ \,
   \Delta \oplus
   \mathfrak{so}(3) \oplus
   \mathfrak{sp}(2,\mathbb{R})
  \, \right] \oplus
  \left( \mathbf{3} , \mathbf{2} \right)^{(+1)} \oplus
  \mathbf{1}^{(+2)}
\end{equation}
where $\Delta$ is the $SO(1,1)$ generator that determines the 5-grading. The generators of $SO(5,2)$ can be realized as nonlinear differential operators acting on a seven-dimensional space $\mathcal{T}$,
 whose coordinates we shall denote as $\mathcal{X}= \left( X^{i,a} , x \right)$, where $X^{i,a}$ transform in the $(3,2)$ representation of $SU(2) \times Sp(2,\mathbb{R})$ subalgebra, with $i=1,2,3$ and $a=1,2$, and $x$ is a singlet coordinate.

There exists a quartic polynomial of the coordinates $X^{i,a}$ 
\begin{equation}
\mathcal{I}_4 (X)
= \eta_{ij} \eta_{kl} \epsilon_{ac} \epsilon_{bd}
  X^{i,a} X^{j,b} X^{k,c} X^{l,d}
\end{equation}
which is an invariant of $SU(2) \times Sp(2,\mathbb{R})$ subgroup, where $\epsilon_{ab}$ is the symplectic invariant tensor of $Sp(2,\mathbb{R})$ and $\eta_{ij}$ is the invariant metric of $SU(2)$ in the adjoint representation, which we choose as $\eta_{ij} = -\delta_{ij}$ to agree with the general conventions of \cite{Gunaydin:2005zz}.

The generators belonging to various grade subspaces will be labelled as follows:
\begin{equation}
\mathfrak{so}(5,2)
= K_{-} \oplus
  U_{i,a} \oplus
  \left[ \Delta \oplus M_{ij} \oplus J_{ab} \right] \oplus
  \widetilde{U}_{i,a} \oplus
  K_{+} 
\end{equation} 
where $M_{ij}$ and $J_{ab}$ are the generators of $SU(2)$ and $Sp(2,\mathbb{R})$ subgroups, respectively. In the nonlinear quasiconformal action of $SO(5,2)$ they are realized as
\begin{equation}
\begin{split}
K_+
&= \frac{1}{2} \left( 2 x^2 - \mathcal{I}_4 \right) \frac{\partial}{\partial x} 
   - \frac{1}{4} \frac{\partial \mathcal{I}_4}{\partial X^{i,a}}
     \eta^{ij} \epsilon^{ab} \frac{\partial}{\partial X^{j,b}}
   + x \, X^{i,a} \frac{\partial}{\partial X^{i,a}}
\\
U_{i,a}
&= \frac{\partial}{\partial X^{i,a}}
   - \eta_{ij} \epsilon_{ab} \, X^{j,b} \frac{\partial}{\partial x}
\\
M_{ij}
&= \eta_{ik} X^{k,a} \frac{\partial}{\partial X^{j,a}}
   - \eta_{jk} X^{k,a} \frac{\partial}{\partial X^{i,a}}
\\
J_{ab}
&= \epsilon_{ac} X^{i,c} \frac{\partial}{\partial X^{i,b}}
   + \epsilon_{bc} X^{i,c} \frac{\partial}{\partial X^{i,a}}
\\
K_-
&= \frac{\partial}{\partial x}
\qquad \qquad
\Delta
= 2 \, x \frac{\partial}{\partial x}
  + X^{i,a} \frac{\partial}{\partial X^{i,a}}
\qquad \qquad
\widetilde{U}_{i,a}
= \commute{U_{i,a}}{K_+}
\end{split}
\end{equation}
where $\epsilon^{ab}$ is the inverse symplectic tensor, such that $\epsilon^{ab} \epsilon_{bc} = {\delta^a}_c$. Substituting the expression for the quartic invariant, one finds the explicit form of the grade +1 generators:
\begin{equation}
\begin{split}
\widetilde{U}_{i,a} 
&= \eta_{ij} \epsilon_{ad} 
   \left(
    \eta_{kl} \epsilon_{bc} X^{j,b} X^{k,c} X^{l,d} 
    - x \, X^{j,d}
   \right)
   \frac{\partial}{\partial x}
   +  x \frac{\partial}{\partial X^{i,a}}
\\
& \quad
   - \eta_{ij} \epsilon_{ab} \, X^{j,b} X^{l,c}
     \frac{\partial}{\partial X^{l,c}}
   - \epsilon_{ad} \eta_{kl} \, X^{l,d} X^{k,c}
     \frac{\partial}{\partial X^{i,c}}
\\ 
& \quad
   + \epsilon_{ad} \eta_{ij} \, X^{l,d} X^{j,b}
     \frac{\partial}{\partial X^{l,b}}
   + \eta_{ij} \epsilon_{bc} X^{j,b} X^{l,c}
     \frac{\partial}{\partial X^{l,a}}
\end{split}
\end{equation}

These $SO(5,2)$ generators satisfy the following commutation relations:
\begin{subequations}
\label{eq:sod4algebra}
\begin{equation}
\begin{split}
\commute{M_{ij}}{M_{kl}}
&= \eta_{jk} M_{il} - \eta_{ik} M_{jl} - \eta_{jl} M_{ik} + \eta_{il} M_{jk}
\\
\commute{J_{ab}}{J_{cd}}
&= \epsilon_{cb} J_{ad} + \epsilon_{ca} J_{bd}
   + \epsilon_{db} J_{ac} + \epsilon_{da} J_{bc}
\end{split}
\end{equation}
\begin{equation}
\begin{split}
\commute{\Delta}{K_\pm}
&= \pm 2 \, K_\pm
\qquad \qquad \qquad
\commute{K_-}{K_+}
= \Delta
\\
\commute{\Delta}{U_{i,a}}
&= - U_{i,a}
\qquad \qquad \qquad \quad
\commute{\Delta}{\widetilde{U}_{i,a}}
= \widetilde{U}_{i,a}
\\
\commute{U_{i,a}}{K_+}
&= \widetilde{U}_{i,a}
\qquad \qquad \qquad \quad
\commute{\widetilde{U}_{i,a}}{K_-}
= - U_{i,a}
\\
\commute{U_{i,a}}{U_{j,b}}
&= 2 \, \eta_{ij} \epsilon_{ab} \, K_-
\qquad \qquad
\commute{\widetilde{U}_{i,a}}{\widetilde{U}_{j,b}}
= 2 \, \eta_{ij} \epsilon_{ab} \, K_+
\end{split}
\end{equation}
\begin{equation}
\begin{split}
\commute{M_{ij}}{U_{k,a}}
&= \eta_{jk} U_{i,a} - \eta_{ik} U_{j,a}
\qquad \qquad
\commute{M_{ij}}{\widetilde{U}_{k,a}}
= \eta_{jk} \widetilde{U}_{i,a} - \eta_{ik} \widetilde{U}_{j,a}
\\
\commute{J_{ab}}{U_{i,c}}
&= \epsilon_{cb} U_{i,a} + \epsilon_{ca} U_{i,b}
\qquad \qquad
\commute{J_{ab}}{\widetilde{U}_{i,c}}
= \epsilon_{cb} \widetilde{U}_{i,a} + \epsilon_{ca} \widetilde{U}_{i,b}
\end{split}
\end{equation}
\begin{equation}
\commute{U_{i,a}}{\widetilde{U}_{j,b}}
= \eta_{ij} \epsilon_{ab} \, \Delta
  - 2 \, \epsilon_{ab} M_{ij}
  - \eta_{ij} J_{ab}
\end{equation}
\end{subequations}
The quartic norm (length) of a vector $\mathcal{X}= \left( X^{i,a} , x \right) \in \mathcal{T}$ is defined as
\begin{equation}
\ell \left( \mathcal{X} \right)
= \mathcal{I}_4 \left( X \right) + 2 \, x^2 \,.
\end{equation}
To see the geometric picture behind the above nonlinear realization, one defines a quartic distance function between any two points $\mathcal{X}$ and $\mathcal{Y}$ in the seven-dimensional space $\mathcal{T}$ as
\begin{equation}
d \left( \mathcal{X} , \mathcal{Y} \right)
= \ell \left( \delta \left( \mathcal{X} , \mathcal{Y} \right) \right)
\end{equation}
where the ``symplectic'' difference $\delta \left( \mathcal{X} , \mathcal{Y} \right)$ is defined as
\begin{equation}
\delta \left( \mathcal{X} , \mathcal{Y} \right)
= \left( X^{i,a} - Y^{i,a} \,,\, x - y - \eta_{ij} \epsilon_{ab} \, X^{i,a} Y^{j,b} \right) \,.
\end{equation}
The lightlike separations between any two points with respect to the quartic distance function are left invariant under the quasiconformal action of $SO(5,2)$. In other words, $SO(5,2)$ acts as the invariance group of a "light-cone" with respect to a quartic distance function in a seven dimensional space.


\section{Minimal unitary representation of $SO(5,2)$}
\label{sec:minrepSO(5,2)}

The quantization of the geometric quasiconformal action of a Lie algebra or a Lie superalgebra leads to its minimal unitary realization. For the case of $SO(5,2)$ this is achieved by splitting the six variables $X^{i,a}$ introduced above into three coordinates $X^i$ and three momenta $P_i$, and introducing a momentum $p$ conjugate to the singlet coordinate $x$: 
\begin{equation}
X^i = X^{i,1}
\qquad \qquad
P_i = \eta_{ij} X^{j,2}
\end{equation}
These coordinates and momenta are then treated as quantum mechanical operators satisfying the canonical commutation relations:
\begin{equation}
\commute{X^i}{P_j} = \delta^i_j
\qquad \qquad
\commute{x}{p} = i
\end{equation}
In the realization that follows, we shall use bosonic oscillators $a_i$ and their hermitian conjugates $a_i^\dag$ defined in terms of $X^i$ and $P_i$ as follows:
\begin{equation}
a_i = \frac{1}{\sqrt{2}} \left( X^i + i \, P_i \right)
\qquad \qquad
a_i^\dag = \frac{1}{\sqrt{2}} \left( X^i - i \, P_i \right)
\end{equation}
They satisfy the commutation relations:
\begin{equation}
\commute{a_i}{a_j^\dag} = \delta_{ij}
\qquad \qquad \qquad
\commute{a_i}{a_j} = \commute{a_i^\dag}{a_j^\dag} = 0
\end{equation}

We shall first give the generators of the minimal unitary realization of $\mathfrak{so}(5,2)$ in the 5-graded basis:
\begin{equation}
\mathfrak{so}(5,2)
= \mathfrak{g}^{(-2)} \oplus \mathfrak{g}^{(-1)} \oplus
  \left[ \,
   \Delta \oplus \mathfrak{su}(2) \oplus \mathfrak{su}(1,1)
  \, \right] \oplus
  \mathfrak{g}^{(+1)} \oplus \mathfrak{g}^{+2)}
\end{equation}
where the generator that defines the 5-grading is simply
\begin{equation}
\Delta = \frac{1}{2} \left( x p + p x \right) \,.
\label{delta}
\end{equation}
The single generator in grade $-2$ subspace is realized in terms of the singlet coordinate $x$ as
\begin{equation}
K_- = \frac{1}{2} x^2
\label{K-}
\end{equation}
and the six generators in grade $-1$ subspace are realized as bilinears of $x$ and the bosonic oscillators $a_i$, $a_i^\dag$ as
\begin{equation}
U_i = x \, a_i
\qquad \qquad \qquad
U_i^\dag = x \, a_i^\dag \,.
\label{Grade-1Bosonic}
\end{equation}
Grade $-2$ and grade $-1$ generators form a Heisenberg subalgebra
\begin{equation}
\commute{U_i}{U_j^\dag} = 2 \, \delta_{ij} \, K_-
\qquad \qquad
\commute{U_i}{U_j} = \commute{U_i^\dag}{U_j^\dag} = 0
\end{equation}
with the generator $K_-$ playing the role of the central charge.
The generators of $\mathfrak{su}(1,1) \subset \mathfrak{g}^{(0)}$ are realized as bilinears of the $a$-type bosonic oscillators as
\begin{equation}
M_+ = \frac{1}{2} a_i^\dag a_i^\dag
\qquad \qquad
M_- = \frac{1}{2} a_i a_i
\qquad \qquad
M_0 = \frac{1}{4} \left( a_i^\dag a_i + a_i a_i^\dag \right)
\label{SU(1,1)M}
\end{equation}
which satisfy the commutation relations
\begin{equation}
\commute{M_-}{M_+} = 2 \, M_0
\qquad \qquad
\commute{M_0}{M_\pm} = \pm \, M_\pm \,.
\end{equation}
We denote this subalgebra as $\mathfrak{su}(1,1)_M$ and its quadratic Casimir
as $\mathcal{M}^2$:
\begin{equation}
\mathcal{C}_2 \left[ \mathfrak{su}(1,1)_M \right]
= \mathcal{M}^2
= {M_0}^2 - \frac{1}{2} \left( M_+ M_- + M_- M_+ \right)
\end{equation}

The $\mathfrak{su}(2)$ subalgebra, denoted as $\mathfrak{su}(2)_L$, of grade 0 subspace is also realized as bilinears of the $a$-type bosonic oscillators as
\begin{equation}
L_i = V^\dag \, \Sigma_i \, V
\end{equation}
where
\begin{equation}
V = \left( \begin{matrix} a_1 \\ a_2 \\ a_3 \end{matrix} \right)
\end{equation}
and $\Sigma_i$ are the $3 \times 3$ adjoint matrices of $SU(2)$ given by
\begin{equation}
\left( \Sigma_i \right)_{jk} = - i \, \epsilon_{ijk} \,.
\end{equation}
They satisfy the commutation relations 
\begin{equation}
\commute{L_i}{L_j} = i \, \epsilon_{ijk} \, L_k \,.
\end{equation}
The quadratic Casimir of $\mathfrak{su}(2)_L$, denoted as $\mathcal{L}^2$, 
\begin{equation}
\mathcal{C}_2 \left[ \mathfrak{su}(2)_L \right]
= \mathcal{L}^2
= {L_1}^2 + {L_2}^2 + {L_3}^2
\end{equation}
is related to that of $\mathfrak{su}(1,1)_M$ as
\begin{equation}
\mathcal{L}^2 = 4 \, \mathcal{M}^2 + \frac{3}{4} \,.
\end{equation}

Upon quantization, the quartic invariant $\mathcal{I}_4$ of $SU(2)_L \times SU(1,1)_M$ goes over to a linear function of the quadratic Casimir of $SU(2)_L \times SU(1,1)_M$. As a consequence one finds that grade $+2$ generator depends only on the quadratic Casimir of $\mathfrak{su}(2)_L$ (or that of $\mathfrak{su}(1,1)_M$) and can be written as follows:
\begin{equation}
\begin{split}
K_+
&= \frac{1}{2} p^2
   + \frac{1}{4 \, x^2} \left( 8 \, \mathcal{M}^2 + \frac{3}{2} \right)
\\
&= \frac{1}{2} p^2
   + \frac{1}{2 \, x^2} \mathcal{L}^2
\end{split}
\end{equation}

Now the six generators in grade $+1$ subspace can be obtained by taking the
commutators between the respective grade $-1$ generators and $K_+$:
\begin{equation}
W_i
= - i \commute{U_i}{K_+}
\qquad \qquad \qquad
W_i^\dag
= - i \commute{U_i^\dag}{K_+}
\label{SO(5,2)grade+1}
\end{equation}
Evaluating the commutators one finds
\begin{equation}
\begin{split}
W_i
&= p \, a_i
   - \frac{i}{x} \left[ a_i + i \, \epsilon_{ijk} \, L_j a_k \right]
\\
W_i^\dag
&= p \, a_i^\dag
   - \frac{i}{x} \left[ a_i^\dag + i \, \epsilon_{ijk} \, L_j a_k^\dag \right]
\end{split}
\end{equation}
Once again, these grade $+2$ and grade $+1$ generators form a Heisenberg algebra
\begin{equation}
\commute{W_i}{W_j^\dag} = 2 \, \delta_{ij} \, K_+
\qquad \qquad
\commute{W_i}{W_j} = \commute{W_i^\dag}{W_j^\dag} = 0
\end{equation}
with the generator $K_+$ playing the role of the central charge.
Grade $-2$ and grade $+1$ generators close into grade $-1$ subspace:
\begin{equation}
\commute{W_i}{K_-} = - i \, U_i
\qquad \qquad
\commute{W_i^\dag}{K_-} = - i \, U_i^\dag
\end{equation}
Grade $\pm 2$ generators, together with the generator $\Delta$ from grade 0 subspace, form a subalgebra $\mathfrak{su}(1,1)$, which we shall denote as $\mathfrak{su}(1,1)_K$:
\begin{equation}
\commute{K_-}{K_+} = i \, \Delta
\qquad \qquad
\commute{\Delta}{K_\pm} = \pm 2 i \, K_\pm
\end{equation}
The quadratic Casimir of  $\mathfrak{su}(1,1)_K$ subalgebra, given by 
\begin{equation}
\mathcal{C}_2 \left[ \mathfrak{su}(1,1)_K \right]
= \mathcal{K}^2
= \Delta^2 - 2 \left( K_+ K_- + K_- K_+ \right)
\end{equation}
is related to the quadratic Casimir of $\mathfrak{su}(2)_L$ (and that of $\mathfrak{su}(1,1)_M$) as
\begin{equation}
\mathcal{K}^2 = - \mathcal{L}^2 + \frac{3}{4} = - 4 \, \mathcal{M}^2 \,.
\end{equation}

The commutators of grade $-1$ (grade $+1$) generators with those of $SU(2)_L \times SU(1,1)_M$ are given below:
\begin{equation}
\begin{aligned}
\commute{M_0}{U_i} &= - \frac{1}{2} \, U_i
\\
\commute{M_+}{U_i} &= - U_i^\dag
\\
\commute{M_-}{U_i} &= 0
\\
\commute{L_i}{U_j} &= i \, \epsilon_{ijk} \, U_k
\end{aligned}
\qquad \qquad \qquad
\begin{aligned}
\commute{M_0}{W_i} &= - \frac{1}{2} \, W_i
\\
\commute{M_+}{W_i} &= - W_i^\dag
\\
\commute{M_-}{W_i} &= 0
\\
\commute{L_i}{W_j} &= i \, \epsilon_{ijk} \, W_k
\end{aligned}
\end{equation}

\begin{equation}
\begin{aligned}
\commute{M_0}{U_i^\dag} &= \frac{1}{2} \, U_i^\dag
\\
\commute{M_+}{U_i^\dag} &= 0
\\
\commute{M_-}{U_i^\dag} &= U_i
\\
\commute{L_i}{U_j^\dag} &= i \, \epsilon_{ijk} \, U_k^\dag
\end{aligned}
\qquad \qquad \qquad
\begin{aligned}
\commute{M_0}{W_i^\dag} &= \frac{1}{2} \, W_i^\dag
\\
\commute{M_+}{W_i^\dag} &= 0
\\
\commute{M_-}{W_i^\dag} &= W_i
\\
\commute{L_i}{W_j^\dag} &= i \, \epsilon_{ijk} \, W_k^\dag
\end{aligned}
\end{equation}
The  commutators between grade $-1$ generators and grade $+1$ generators can be written in terms of grade 0 generators as follows:
\begin{equation}
\begin{aligned}
\commute{U_i}{W_j}
&= 2 i \, \delta_{ij} \, M_-
\\
\commute{U_i^\dag}{W_j^\dag}
&= 2 i \, \delta_{ij} \, M_+
\end{aligned}
\qquad \qquad
\begin{aligned}
\commute{U_i^\dag}{W_j}
&= 2 i \, \delta_{ij} \, M_0
   - 2 \, \epsilon_{ijk} \, L_k
   - \delta_{ij} \, \Delta
\\
\commute{U_i}{W_j^\dag}
&= 2 i \, \delta_{ij} \, M_0
   + 2 \, \epsilon_{ijk} \, L_k
   + \delta_{ij} \, \Delta
\end{aligned}
\end{equation}

Finally, we present the quadratic Casimir of $\mathfrak{so}(5,2)$. Noting that the following combination of bilinears, formed in terms of the generators in grade $\pm1$ subspaces, reduces to the quadratic Casimir of $\mathfrak{su}(1,1)_K$ modulo some additive and multiplicative constants,
\begin{equation}
\left[ U W \right]
= U_i W_i^\dag + W_i^\dag U_i - U_i^\dag W_i - W_i U_i^\dag
= - 4 i \, \mathcal{K}^2 + 12 i
\end{equation}
one can show that there exists a two-parameter family of degree two polynomials of $\mathfrak{so}(5,2)$ generators that reduces to a $c$-number for the minimal unitary realization, according to Joseph's theorem \cite{Joseph:1974,joseph1976minimal}:
\begin{equation}
\begin{split}
&\mathcal{C}_2 \left[ \mathfrak{su}(2)_L \right]
+ k_1 \, \mathcal{C}_2 \left[ \mathfrak{su}(1,1)_M \right]
+ k_2 \, \mathcal{C}_2 \left[ \mathfrak{su}(1,1)_K \right]
+ \frac{i}{4} \left( 1 + \frac{k_1}{4} - k_2 \right)
  \left[ UW \right]
\\
& \qquad
= - \frac{3}{4} \, k_1 + 3 \, k_2 - \frac{9}{4}
\end{split}
\end{equation}
The quadratic Casimir of $\mathfrak{so}(5,2)$ corresponds to $k_1 = 2$ and $k_2 = - \frac{1}{2}$:
\begin{equation}
\mathcal{C}_2 \left[ \mathfrak{so}(5,2) \right] = - \frac{21}{4}
\end{equation}


\section{Noncompact 3-grading of $\mathfrak{so}(5,2)$ with respect to the subalgebra $\mathfrak{so}(4,1) \oplus \mathfrak{so}(1,1) \approx \mathfrak{usp}(2,2) \oplus \mathfrak{so}(1,1)$}
\label{sec:SO(5,2)nc3G}

We should note that when one goes to the covering group $Spin(5,2)$ of $SO(5,2)$, the subgroups $SO(5)$ and $SO(4,1)$ go over to their covering groups $U\!Sp(4)$ and $U\!Sp(2,2)$, respectively. Considered as the five-dimensional conformal group, $SO(5,2)$ has a natural 3-grading defined by the generator $\mathcal{D}$ of dilatations whose eigenvalues determine the conformal dimensions of operators and states. Let us denote the corresponding 3-graded decomposition of $\mathfrak{so}(5,2)$ as
\begin{equation}
\mathfrak{so}(5,2)
= \mathfrak{N}^- \oplus
  \mathfrak{N}^0 \oplus
  \mathfrak{N}^+
\end{equation}
where $\mathfrak{N}^0 = \mathfrak{so}(4,1) \oplus \mathfrak{so}(1,1)_{\mathcal{D}}$, with the subalgebra $\mathfrak{so}(4,1)$ in $\mathfrak{N}^{0}$ representing the Lorentz algebra in five dimensions. The \emph{noncompact} dilatation generator $\mathfrak{so}(1,1)_{\mathcal{D}}$ is given by
\begin{equation}
\mathcal{D}
= \frac{1}{2} \left[ \Delta - i \left( M_+ - M_- \right) \right]
\label{Dilatation}
\end{equation}
and the generators belonging to $\mathfrak{N}^{\pm}$ and $\mathfrak{N}^0$ subspaces are as follows:
\begin{equation}
\begin{split}
\mathfrak{N}^-
&= K_- \oplus
   \left[ M_0 - \frac{1}{2} \left( M_+ + M_- \right) \right]  \oplus
   \left( U_i - U_i^\dag \right)
\\
\mathfrak{N}^0
&= \mathcal{D} \oplus
   \frac{1}{2} \left[ \Delta + i \left( M_+ - M_- \right) \right] \oplus
   L_i \oplus
   \left( U_i + U_i^\dag \right) \oplus
   \left( W_i - W_i^\dag \right)
\\
\mathfrak{N}^+
&= K_+ \oplus
   \left[ M_0 + \frac{1}{2} \left( M_+ + M_- \right) \right]  \oplus
   \left( W_i + W_i^\dag \right)
\end{split}
\end{equation}
The Lorentz group generators $\mathcal{M}_{\mu\nu}$ ($\mu,\nu = 0,1,2,3,4$) are given by

\begin{equation} \label{lorentz}
\begin{split}
\mathcal{M}_{0i}
= \frac{1}{2\sqrt{2}} \left( U_i + U_i^\dag \right)
  + \frac{i}{2\sqrt{2}} \left( W_i - W_i^\dag \right)
& \qquad \qquad
\mathcal{M}_{ij}
= - \epsilon_{ijk} \, L_k
\\
\mathcal{M}_{i4}
= \frac{1}{2\sqrt{2}} \left( U_i + U_i^\dag \right)
  - \frac{i}{2\sqrt{2}} \left( W_i - W_i^\dag \right)
& \qquad \qquad
\mathcal{M}_{04}
= \frac{1}{2} \left[ \Delta + i \left( M_+ - M_- \right) \right]
\end{split}
\end{equation}
and satisfy the commutation relations of $\mathfrak{so}(4,1)$:
\begin{equation}
\commute{\mathcal{M}_{\mu\nu}}{\mathcal{M}_{\rho\tau}}
= i \left(
     \eta_{\nu\rho} \mathcal{M}_{\mu\tau}
     - \eta_{\mu\rho} \mathcal{M}_{\nu\tau}
     - \eta_{\nu\tau} \mathcal{M}_{\mu\rho}
     + \eta_{\mu\tau} \mathcal{M}_{\nu\rho}
    \right)
\end{equation}
where $\eta_{\mu\nu} = \mathrm{diag} (-,+,+,+,+)$. 

The rotation group $SO(4)$ splits into two $SU(2)$ subgroups, denoted by $SU(2)_A$ and $SU(2)_{\mathring{A}}$, whose generators are given by
\begin{equation}
A_i = - \frac{1}{4} \left( \epsilon_{ijk} \mathcal{M}_{jk} - 2 \, \mathcal{M}_{i4} \right)
\qquad \qquad
\mathring{A}_i = - \frac{1}{4} \left( \epsilon_{ijk} \mathcal{M}_{jk} + 2 \, \mathcal{M}_{i4} \right)
\label{LeftRightSU(2)}
\end{equation}
satisfying the commutation relations
\begin{equation}
\commute{A_i}{A_j} = i \, \epsilon_{ijk} A_k
\qquad \qquad
\commute{\mathring{A}_i}{\mathring{A}_j} = i \, \epsilon_{ijk} \mathring{A}_k
\end{equation}

The translation generators $\mathcal{P}_\mu$ ($\mu = 0,1,2,3,4$) of the conformal group $SO(5,2)$ are given by
\begin{equation}
\begin{split}
\mathcal{P}_0
&= K_+ + M_0 + \frac{1}{2} \left( M_+ + M_- \right)
\\
\mathcal{P}_i
&= \frac{1}{\sqrt{2}} \left( W_i + W_i^\dag \right) \qquad \quad (i = 1,2,3)
\\
\mathcal{P}_4
&= K_+ - M_0 - \frac{1}{2} \left( M_+ + M_- \right)
\end{split}
\end{equation}
and the special conformal generators $\mathcal{K}_\mu$ ($\mu = 0,1,2,3,4$) are given by
\begin{equation}
\begin{split}
\mathcal{K}_0
&= K_- + M_0 - \frac{1}{2} \left( M_+ + M_- \right)
\\
\mathcal{K}_i
&= - \frac{i}{\sqrt{2}} \left( U_i - U_i^\dag \right) \qquad \quad (i = 1,2,3)
\\
\mathcal{K}_4
&= - K_- + M_0 - \frac{1}{2} \left( M_+ + M_- \right) \,.
\end{split}
\end{equation}
These generators satisfy the commutation relations of $SO(5,2)$ as the five-dimensional conformal algebra:
\begin{equation}
\begin{split}
\commute{\mathcal{M}_{\mu\nu}}{\mathcal{M}_{\rho\tau}}
&= i \left(
     \eta_{\nu\rho} \mathcal{M}_{\mu\tau}
     - \eta_{\mu\rho} \mathcal{M}_{\nu\tau}
     - \eta_{\nu\tau} \mathcal{M}_{\mu\rho}
     + \eta_{\mu\tau} \mathcal{M}_{\nu\rho}
    \right)
\\
\commute{\mathcal{P}_\mu}{\mathcal{M}_{\nu\rho}}
&= i \left( \eta_{\mu\nu} \, \mathcal{P}_\rho
            - \eta_{\mu\rho} \, \mathcal{P}_\nu
     \right)
\\
\commute{\mathcal{K}_\mu}{\mathcal{M}_{\nu\rho}}
&= i \left( \eta_{\mu\nu} \, \mathcal{K}_\rho
            - \eta_{\mu\rho} \, \mathcal{K}_\nu
     \right)
\\
\commute{\mathcal{D}}{\mathcal{M}_{\mu\nu}}
&= \commute{\mathcal{P}_\mu}{\mathcal{P}_\nu}
 = \commute{\mathcal{K}_\mu}{\mathcal{K}_\nu}
 = 0
\\
\commute{\mathcal{D}}{\mathcal{P}_\mu}
&= + i \, \mathcal{P}_\mu
\qquad \qquad
\commute{\mathcal{D}}{\mathcal{K}_\mu}
 = - i \, \mathcal{K}_\mu
\\
\commute{\mathcal{P}_\mu}{\mathcal{K}_\nu}
&= 2 i \left( \eta_{\mu\nu} \, \mathcal{D} + \mathcal{M}_{\mu\nu} \right)
\end{split}
\end{equation}

We should note that the Poincar\'e mass operator in five dimensions vanishes identically
\begin{equation}
\mathcal{M}^2 = \eta_{\mu\nu} \mathcal{P}^\mu\mathcal{P}^\nu
              = 0
\end{equation}
for the minimal unitary realization.

In appendix \ref{App:SO(5,2)}, we present the relation between the generators $M_{AB}$ ($A,B = 0,\dots,6$) of $SO(5,2)$ and the generators in the noncompact three-grading.


\section{Compact 3-grading of $\mathfrak{so}(5,2)$ with respect to the subalgebra $\mathfrak{so}(5) \oplus \mathfrak{so}(2) \approx \mathfrak{usp}(4) \oplus  \mathfrak{u}(1)$}
\label{sec:SO(5,2)c3G}

The Lie algebra $\mathfrak{so}(5,2)$ has a 3-grading, with respect to its maximal compact subalgebra $\mathfrak{C}^0 = \mathfrak{so}(5) \oplus \mathfrak{so}(2) = \mathfrak{usp}(4) \oplus \mathfrak{u}(1)$, determined by the $\mathfrak{u}(1)$ generator
\begin{equation}
H = \frac{1}{2} \left( K_+ + K_- \right) + M_0
\label{SO(2)generator}
\end{equation}
such that
\begin{equation}
\mathfrak{so}(5,2)
= \mathfrak{C}^- \oplus
  \mathfrak{C}^0 \oplus
   \mathfrak{C}^+
\end{equation}
and  satisfy
\begin{equation}
\commute{H}{\mathfrak{C}^+} = + \, \mathfrak{C}^+
\qquad \qquad \qquad
\commute{H}{\mathfrak{C}^-} = - \, \mathfrak{C}^- \,.
\end{equation}

In this decomposition, the generators belonging to $\mathfrak{C}^{\pm}$ and $\mathfrak{C}^0$ subspaces are as follows:
\begin{equation}
\begin{split}
\mathfrak{C}^0
 &= \left[
     L_i \oplus
     \left( \frac{1}{2} \left( K_+ + K_- \right) - M_0 \right) \oplus
     \left( U_i - i \, W_i \right) \oplus
     \left( U_i^\dag + i \, W_i^\dag \right)
    \right]
\\
 & \qquad
   \oplus
   \left( \frac{1}{2} \left( K_+ + K_- \right) + M_0 \right)
\\
\mathfrak{C}^+
 &= \left( U_i^\dag - i \, W_i^\dag \right) \oplus
    M_+ \oplus
    \left[ \Delta - i \left( K_+ - K_- \right) \right]
\\
\mathfrak{C}^-
 &= \left( U_i + i \, W_i \right) \oplus
    M_- \oplus
    \left[ \Delta + i \left( K_+ - K_- \right) \right]
\end{split}
\end{equation}
In the above 3-grading, the operators that belong to $\mathfrak{C}^+$ subspace are the Hermitian conjugates of the operators that belong to $\mathfrak{C}^-$ subspace. In the corresponding minimal unitary realization, one takes only the hermitian linear combinations of these operators as generators of $\mathfrak{so}(5,2)$. The generator $H$ is the conformal Hamiltonian or the $AdS$ energy, depending on whether one is considering $SO(5,2)$ as the five-dimensional conformal group or as the six-dimensional $AdS$ group. We shall refer to this grading as the \emph{compact} 3-grading. We should also note that, in the earlier noncompact 3-grading of $\mathfrak{so}(5,2)$ with respect to $\mathfrak{N}^0 = \mathfrak{so}(4,1) \oplus \mathfrak{so}(1,1)_{\mathcal{D}}$, this $AdS$ energy corresponds to $\frac{1}{2} \left( \mathcal{P}_0 + \mathcal{K}_0 \right)$.


\subsection{$SO(5)$-covariant basis}
\label{subsec:SO(5,2)inSO(5)cov}

The $\mathfrak{so}(5)$ generators $\widetilde{M}_{MN}$ ($M,N = 1,2,3,4,5$) in grade zero subspace $\mathfrak{C}^0$ are given by
\begin{equation}
\begin{aligned}
\widetilde{M}_{ij}
&= - \epsilon_{ijk} \, L_k
\\
\widetilde{M}_{45}
&= \frac{1}{2} \left( K_+ + K_- \right) - M_0
\end{aligned}
\qquad \qquad
\begin{aligned}
\widetilde{M}_{i4}
&= \frac{1}{2\sqrt{2}} \left( U_i - i \, W_i \right)
   + \frac{1}{2\sqrt{2}} \left( U_i^\dag + i \, W_i^\dag \right)
\\
\widetilde{M}_{i5}
&= \frac{i}{2\sqrt{2}} \left( U_i - i \, W_i \right)
   - \frac{i}{2\sqrt{2}} \left( U_i^\dag + i \, W_i^\dag \right)
\end{aligned}
\label{SO(5)generators}
\end{equation}
and satisfy the commutation relations
\begin{equation}
\commute{\widetilde{M}_{MN}}{\widetilde{M}_{PQ}}
= i \left(
     \delta_{NP} \widetilde{M}_{MQ} - \delta_{MP} \widetilde{M}_{NQ}
     - \delta_{NQ} \widetilde{M}_{MP} + \delta_{MQ} \widetilde{M}_{NP}
    \right) \,.
\end{equation}
Clearly the generators $\widetilde{M}_{ij} \oplus \widetilde{M}_{45}$ form the $\mathfrak{so}(3)_L \oplus \mathfrak{so}(2)_Y \approx \mathfrak{su}(2)_L \oplus \mathfrak{u}(1)_Y$ subalgebra of $\mathfrak{so}(5) \approx \mathfrak{usp}(4)$. 
We shall label the five operators that belong to grade $+1$ subspace $\mathfrak{C}^+$ as $\widetilde{B}_M^\dag$ ($M = 1,2,3,4,5$) where 
\begin{equation}
\begin{split}
\widetilde{B}_i^\dag
&= \frac{1}{\sqrt{2}} \left( U_i^\dag - i \, W_i^\dag \right)
\qquad \qquad (i = 1,2,3)
\\
\widetilde{B}_4^\dag
&= \frac{1}{2} \left[ \Delta - i \left( K_+ - K_- \right) \right] + i \, M_+
\\
\widetilde{B}_5^\dag
&= \frac{i}{2} \left[ \Delta - i \left( K_+ - K_- \right) \right] + M_+
\end{split}
\label{SO(5,2)3gradingC+}
\end{equation}

These $\mathfrak{C}^+$ operators satisfy the following important relation:
\begin{equation}
\widetilde{B}_M^\dag \widetilde{B}_M^\dag
= \widetilde{B}_1^\dag \widetilde{B}_1^\dag
  + \widetilde{B}_2^\dag \widetilde{B}_2^\dag
  + \dots
  + \widetilde{B}_5^\dag \widetilde{B}_5^\dag
= 0
\label{MinrepC+Constraint}
\end{equation}
which corresponds to the masslessness condition in the noncompact picture.
We shall label the five operators that belong to grade $-1$ subspace $\mathfrak{C}^-$, which are the hermitian conjugates of those in $\mathfrak{C}^+$, as  $\widetilde{B}_M$ ($M = 1,2,3,4,5$) where
\begin{equation}
\begin{split}
\widetilde{B}_i
&= \frac{1}{\sqrt{2}} \left( U_i + i \, W_i \right)
\qquad \qquad (i = 1,2,3)
\\
\widetilde{B}_4
&= \frac{1}{2} \left[ \Delta + i \left( K_+ - K_- \right) \right] - i \, M_-
\\
\widetilde{B}_5
&= - \frac{i}{2} \left[ \Delta + i \left( K_+ - K_- \right) \right] + M_- \,.
\end{split}
\label{SO(5,2)3gradingC-}
\end{equation}

The commutation relations of the $SO(5,2)$ generators in this compact basis are:
\begin{equation}
\begin{split}
\commute{\widetilde{M}_{MN}}{\widetilde{M}_{PQ}}
&= i \left(
      \delta_{NP} \widetilde{M}_{MQ} - \delta_{MP} \widetilde{M}_{NQ}
      - \delta_{NQ} \widetilde{M}_{MP} + \delta_{MQ} \widetilde{M}_{NP}
     \right)
\\
\commute{\widetilde{B}_M^\dag}{\widetilde{M}_{NP}}
&= i \left(
      \delta_{MN} \, \widetilde{B}_P^\dag - \delta_{MP} \, \widetilde{B}_N^\dag
     \right)
\\
\commute{\widetilde{B}_M}{\widetilde{M}_{NP}}
&= i \left(
      \delta_{MN} \, \widetilde{B}_P - \delta_{MP} \, \widetilde{B}_N
     \right)
\\
\commute{H}{\widetilde{M}_{MN}}
&= \commute{\widetilde{B}_M^\dag}{\widetilde{B}_N^\dag}
 = \commute{\widetilde{B}_M}{\widetilde{B}_N}
 = 0
\\
\commute{H}{\widetilde{B}_M^\dag}
&= + \widetilde{B}_M^\dag
\qquad \qquad
\commute{H}{\widetilde{B}_M}
 = - \widetilde{B}_M
\\
\commute{\widetilde{B}_M^\dag}{\widetilde{B}_N}
&= 2  \left( - \delta_{MN} \, H + i \, \widetilde{M}_{MN} \right)
\end{split}
\end{equation}


\subsection{$U\!Sp(4)$-covariant basis}
\label{subsec:SO(5,2)inUSP(4)cov}

In this subsection we shall reformulate the  compact 3-grading of the Lie algebra $\mathfrak{so}(5,2)$ in the $U\!Sp(4)$-covariant form. As gamma-matrices $\left( \gamma_M \right)^I_{~J}$ in the five-dimensional Euclidean space we choose the following hermitian matrices:
\begin{equation}
\begin{split}
&\left( \gamma_1 \right)^I_{~J}
= \sigma_2 \otimes \sigma_1
= \left(
   \begin{matrix}
    0 & 0 & 0 & -i \\ 0 & 0 & -i & 0 \\ 0 & i & 0 & 0 \\ i & 0 & 0 & 0
   \end{matrix}
  \right)
\qquad \qquad
\left( \gamma_2 \right)^I_{~J}
= \sigma_2 \otimes \sigma_2
= \left(
   \begin{matrix}
    0 & 0 & 0 & -1 \\ 0 & 0 & 1 & 0 \\ 0 & 1 & 0 & 0 \\ -1 & 0 & 0 & 0
   \end{matrix}
  \right)
\\
&\left( \gamma_3 \right)^I_{~J}
= - \sigma_2 \otimes \sigma_3
= \left(
   \begin{matrix}
    0 & 0 & i & 0 \\ 0 & 0 & 0 & -i \\ -i & 0 & 0 & 0 \\ 0 & i & 0 & 0
   \end{matrix}
  \right)
\qquad \quad
\left( \gamma_4 \right)^I_{~J}
= - \sigma_1 \otimes \mathbb{I}_2
= \left(
   \begin{matrix}
    0 & 0 & -1 & 0 \\ 0 & 0 & 0 & -1 \\ -1 & 0 & 0 & 0 \\ 0 & -1 & 0 & 0
   \end{matrix}
  \right)
\\
&\left( \gamma_5 \right)^I_{~J}
= \sigma_3 \otimes \mathbb{I}_2
= \left(
   \begin{matrix}
    1 & 0 & 0 & 0 \\ 0 & 1 & 0 & 0 \\ 0 & 0 & -1 & 0 \\ 0 & 0 & 0 & -1
   \end{matrix}
  \right)
\end{split}
\label{SO(5)gamma}
\end{equation}
which satisfy \begin{equation}
\anticommute{\gamma_M}{\gamma_N} = 2 \, \delta_{MN} \, \mathbb{I}_4
\end{equation}
where $\mathbb{I}_n$ are the $n \times n$ identity matrices. As the charge conjugation matrix, we choose the antisymmetric matrix
\begin{equation}
\left( \mathcal{C}_5 \right)_{IJ}
= \left( \mathcal{C}_5 \right)^{IJ}
= \mathbb{I}_2 \otimes i \sigma_2
= \left(
   \begin{matrix}
    0 & 1 & 0 & 0 \\ -1 & 0 & 0 & 0 \\ 0 & 0 & 0 & 1 \\ 0 & 0 & -1 & 0
   \end{matrix}
  \right)
\end{equation}
where $I,J = 1,2,3,4$ are the spinor indices of the covering group $U\!Sp(4)$ of $SO(5)$. It can be identified with the symplectic metric $\Omega_{IJ} = \Omega^{IJ}$ of $U\!Sp(4)$ and will be used to raise and lower spinorial indices. We find that $\left( \mathcal{C}_5 \gamma_M \right)_{IJ}$  are antisymmetric and $\left( \mathcal{C}_5 \left[ \gamma_M , \gamma_N \right] \right)_{IJ}$  are symmetric with respect to the indices $I$ and $J$.
Using them we can convert the vector indices $M,N,\dots$ of $SO(5)$ into bispinorial indices $[IJ],\dots$ of $USp(4)$. 
We find
\begin{equation}
\begin{split}
\mathcal{B}_{IJ}
&= \left( \mathcal{C}_5 \gamma_M \right)_{IJ} \widetilde{B}_M
 \in \, \mathfrak{C}^-
\\
\overline{\mathcal{B}}_{IJ}
&= \Omega_{IK} \mathcal{B}_{KL}^\dag \Omega_{LJ}
 = - \left( \mathcal{C}_5 \gamma_M \right)_{IJ} \widetilde{B}_M^\dag
 \in \,\mathfrak{C}^+ \,.
\end{split}
\end{equation}
Note that they satisfy the symplectic traceless conditions
\begin{equation}
\Omega_{IJ} \mathcal{B}_{IJ} = 0
\qquad \qquad \qquad
\Omega_{IJ} \overline{\mathcal{B}}_{IJ} = 0 \,.
\end{equation}

The generators in grade 0 subspace that form the subalgebra $\mathfrak{usp}(4)$ are realized as
\begin{equation}
U_{IJ}
= \frac{i}{4} \left( \mathcal{C}_5 \left[ \gamma_M , \gamma_N \right] \right)_{IJ}
  \widetilde{M}_{MN}
\end{equation}
where $\widetilde{M}_{MN}$ are the $\mathfrak{so}(5)$ generators given in equation (\ref{SO(5)generators}).

They satisfy \begin{equation}
U_{IJ} = \Omega_{IK} U_{KL}^\dag \Omega_{LJ} \,.
\end{equation}

The commutation relations of these $SO(5,2)$ generators in this $U\!Sp(4)$-covariant compact basis have the following form:
\begin{equation}
\begin{split}
\commute{U_{IJ}}{U_{KL}}
&= \Omega_{JK} \, U_{IL} + \Omega_{IK} \, U_{JL}
   + \Omega_{JL} \, U_{IK} + \Omega_{IL} \, U_{JK}
\\
\commute{U_{IJ}}{\overline{\mathcal{B}}_{KL}}
&= \Omega_{JK} \, \overline{\mathcal{B}}_{IL}
   + \Omega_{IK} \, \overline{\mathcal{B}}_{JL}
   - \Omega_{JL} \, \overline{\mathcal{B}}_{IK}
   - \Omega_{IL} \, \overline{\mathcal{B}}_{JK}
\\
\commute{U_{IJ}}{\mathcal{B}_{KL}}
&= \Omega_{JK} \, \mathcal{B}_{IL} + \Omega_{IK} \, \mathcal{B}_{JL}
   - \Omega_{JL} \, \mathcal{B}_{IK} - \Omega_{IL} \, \mathcal{B}_{JK}
\\
\commute{H}{U_{IJ}}
&= \commute{\overline{\mathcal{B}}_{IJ}}{\overline{\mathcal{B}}_{KL}}
 = \commute{\mathcal{B}_{IJ}}{\mathcal{B}_{KL}}
 = 0
\\
\commute{H}{\overline{\mathcal{B}}_{IJ}}
&= + \overline{\mathcal{B}}_{IJ}
\qquad \qquad
\commute{H}{\mathcal{B}_{IJ}}
 = - \mathcal{B}_{IJ}
\\
\end{split}
\end{equation}

The constraint on grade $+1$ operators given in equation (\ref{MinrepC+Constraint}) becomes \begin{equation}
\Omega_{IK} \Omega_{JL} \overline{\mathcal{B}}_{IJ} \overline{\mathcal{B}}_{KL}
= 0
\end{equation}
in the $U\!Sp(4)$-covariant basis.


\section{Distinguished $SU(1,1)_K$ subgroup of $SO(5,2)$ generated by the isotonic (singular) oscillators}
\label{sec:SU(1,1)_K}

Note that in terms of the oscillators $a_i$ (and their respective hermitian conjugates $a_i^\dag$) and the singlet coordinate $x$ and its conjugate momentum $p$, the $\mathfrak{u}(1)$ generator $H$, as given in equation (\ref{SO(2)generator}), has the following form:
\begin{equation}
\begin{split}
H = \frac{1}{2} \left( K_+ + K_- \right) + M_0
 &= \frac{1}{4} \left( x^2 + p^2 \right)
     + \frac{1}{4 \, x^2} \mathcal{L}^2
     + \frac{1}{2} \, a_i^\dag a_i
     + \frac{3}{4}
\\
  &= H_\odot + H_a
\end{split}
\end{equation}
where
\begin{equation}
H_\odot
= \frac{1}{4} \left( x^2 + p^2 \right) + \frac{1}{4 \, x^2} \mathcal{L}^2
\qquad \qquad
H_a
= \frac{1}{2} \, a_i^\dag a_i + \frac{3}{4} \,.
\end{equation}
This $\mathfrak{u}(1)$ generator $H$ is the six-dimensional $AdS$ energy operator or the five-dimensional conformal Hamiltonian. $H_a$ ($= M_0$) is the contribution to the Hamiltonian from the $a$-type standard non-singular bosonic oscillators. On the other hand, $H_\odot$ is the Hamiltonian of a singular harmonic oscillator with a potential function
\begin{equation}
V \left( x \right) = \frac{\mathcal{G}}{x^2}
\end{equation}
where
\begin{equation}
\mathcal{G} = \frac{1}{2} \mathcal{L}^2 \,.
\label{IsotonicCouplingConstant}
\end{equation}
This $H_\odot$ is exactly of the form of the Hamiltonian of conformal quantum mechanics \cite{deAlfaro:1976je} with $\mathcal{G}$ playing the role of the ``coupling constant'' \cite{Gunaydin:2001bt}. In some literature it is referred to as the isotonic oscillator \cite{Casahorran:1995vt,carinena-2007}. It is also of the form that appears in the Calogero models in \cite{Calogero:1969af,Calogero:1970nt}.

Let us now consider this singular harmonic oscillator Hamiltonian
\begin{equation}
\begin{split}
H_\odot
 = \frac{1}{2} \left( K_+ + K_- \right)
&= \frac{1}{4} \left( x^2 + p^2 \right)
   + \frac{1}{2 \, x^2} \mathcal{G}
\\
&= \frac{1}{4} \left( x^2 - \frac{\partial^2}{\partial x^2} \right)
   + \frac{1}{2 \, x^2} \mathcal{G} \,.
\end{split}
\label{SingularHamiltonian}
\end{equation}

The following two linear combinations of the operators $\widetilde{B}_4^\dag$ , $\widetilde{B}_5^\dag$ from the $\mathfrak{C}^+$ subspace and $\widetilde{B}_4$ , $\widetilde{B}_5$ from the $\mathfrak{C}^-$ subspace of $\mathfrak{so}(5,2)$:
\begin{equation}
\begin{split}
\widetilde{B}_\odot^\dag
 = - \frac{i}{2} \left( \widetilde{B}_4^\dag - i \, \widetilde{B}_5^\dag \right)
 = - \frac{i}{2} \left[ \Delta - i \left( K_+ - K_- \right) \right]
&= \frac{1}{4} \left( x - i p \right)^2
   - \frac{1}{2 \, x^2} \mathcal{G}
\\
 &= \frac{1}{4} \left( x - \frac{\partial}{\partial x} \right)^2
    - \frac{1}{2 \, x^2} \mathcal{G}
\\
\widetilde{B}_\odot
 = \frac{i}{2} \left( \widetilde{B}_4 + i \, \widetilde{B}_5 \right)
 = \frac{i}{2} \left[ \Delta + i \left( K_+ - K_- \right) \right]
&= \frac{1}{4} \left( x + i p \right)^2
   - \frac{1}{2 \, x^2} \mathcal{G}
\\
&= \frac{1}{4} \left( x + \frac{\partial}{\partial x} \right)^2
   - \frac{1}{2 \, x^2} \mathcal{G}
\end{split}
\end{equation}
close into $H_\odot$ under commutation, and they generate the distinguished  $\mathfrak{su}(1,1)_K$ subalgebra:\footnote{This is the $SU(1,1)$ subgroup generated by the longest root vector.}
\begin{equation}
\commute{\widetilde{B}_\odot}{\widetilde{B}_\odot^\dag} = 2 \, H_\odot
\qquad \qquad
\commute{H_\odot}{\widetilde{B}_\odot^\dag} = + \, \widetilde{B}_\odot^\dag
\qquad \qquad
\commute{H_\odot}{\widetilde{B}_\odot} = - \, \widetilde{B}_\odot
\end{equation}
For the positive energy unitary representations of $SO(5,2)$, the relevant unitary realizations of $SU(1,1)_K$ are also of the positive energy type.

Now consider the Fock space of the $a$-type oscillators, whose vacuum state $\ket{0}$ is annihilated by all $a_i$:
\begin{equation}
a_i \ket{0} = 0
\qquad \qquad
\left( i = 1,2,3 \right)
\end{equation}
A ``particle basis'' in this Fock space is provided by the states of the form
\begin{equation}
\ket{n_1,n_2,n_3}
= \prod_i \frac{1}{\sqrt{n_{i} !}} \, ( \, a_i^\dag \, )^{n_{i}} \ket{0}
\end{equation}
where $n_{i}$ are non-negative integers.

For a given eigenvalue $g$ of the operator $\mathcal{G}$ (as given in equation (\ref{IsotonicCouplingConstant})), the state(s) corresponding to the lowest energy eigenvalue of $H_\odot$ are the superpositions of tensor product states of the form $\psi_0^{(\alpha_g)} \left( x \right) \ket{\Lambda_g}$, where $\ket{\Lambda_g}$, independent of $x$, is an eigenstate of $\mathcal{G}$ with eigenvalue $g$ in the Fock space of $a$-type oscillators, and $\psi_0^{(\alpha_g)} \left( x \right)$ is a function that satisfies
\begin{equation}
\widetilde{B}_\odot \, \psi_0^{(\alpha_g)} \left( x \right) \ket{\Lambda_g}
= 0 \,.
\end{equation}
Such solutions are given by \cite{MR858831}
\begin{equation}
\psi_0^{(\alpha_g)} \left( x \right)
= C_0 \, x^{\alpha_g} e^{-x^2/2}
\label{singularwavefunctions}
\end{equation}
where $C_0$ is a normalization constant and
\begin{equation}
\alpha_g
= \frac{1}{2} \pm \sqrt{2 g + \frac{1}{4}} \,.
\end{equation}
The Hermiticity of $H_\odot$ implies that
\begin{equation}
g \geq - \frac{1}{8}
\end{equation}
and the normalizability of the state given in equation (\ref{singularwavefunctions}) imposes the constraint
\begin{equation}
\alpha_g > - \frac{1}{2} \,.
\end{equation}

Clearly, $\psi_0^{(\alpha_g)} \left( x \right) \ket{\Lambda_g}$ is an eigenstate of $H_\odot$
with eigenvalue
\begin{equation}
E_{\odot,0}^{(\alpha_g)}
= \frac{\alpha_g}{2} + \frac{1}{4}
= \frac{1}{2} \pm \frac{1}{2} \sqrt{2 g + \frac{1}{4}} \,.
\label{alpha_g}
\end{equation}

For the minrep of $SO(5,2)$ given earlier, the lowest possible value of $g$ is zero, \emph{i.e.} when $\ket{\Lambda_g}$ is simply the Fock vacuum $\ket{0}$ of bosonic oscillators $a_i$ ($i=1,2,3$). For $g = 0$ we have  two possible values of $\alpha_g$, namely 0 and 1. It turns out that, even though the state with $\alpha_g = 0$ has lower energy than that with $\alpha_g = 1$, it leads to non-normalizable states under the action of $SO(5)$ when we extend $SU(1,1)_K$ to $SO(5,2)$. Therefore, we choose the state 
\begin{equation}
\psi_0^{(\alpha_g=1)} \left( x \right) \, \ket{0}
= C_0 \, x \, e^{-x^2/2} \ket{0}
\end{equation}
with energy
\begin{equation}
E_{\odot,0}^{(\alpha_g=1)} = \frac{3}{4}
\end{equation}
as the lowest energy ``ground state'' of $H_\odot$. The higher energy eigenstates of $H_\odot$ can be obtained from this ground state by acting on it repeatedly with the raising operator $\widetilde{B}_\odot^\dag$:
\begin{equation}
\psi_n^{(\alpha_g=1)} \left( x \right) \, \ket{0}
 = C_n \, ( \widetilde{B}_\odot^\dag )^n \,
   \psi_0^{(\alpha_g=1)} \left( x \right) \ket{0}
\label{isotonicgroundstate}
\end{equation}
where $C_n$ are normalization constants. They correspond to energy eigenvalues
\begin{equation}
E_{\odot,n}^{(\alpha_g=1)}
 = E_{\odot,0}^{(\alpha_g=1)} + n
 = \frac{3}{4} + n \,.
\end{equation}
We shall denote the corresponding states as 
\begin{equation*}
| \, \psi^{(\alpha_g)}_n \, \rangle \qquad \qquad (n=0,1,2,...)
\end{equation*}
which form the particle basis of the unitary irreducible representation of $SU(1,1)_K$ with the lowest weight vector $|\psi^{(\alpha_g)}_0 \rangle $.


\section{$K$-type decomposition of the minimal unitary representation  of $SO(5,2)$}
\label{sec:undeformedminrep}

In this section we shall give the decomposition of the Hilbert space of the minimal unitary representation of $SO(5,2)$ with respect to its  maximal compact subgroup $SO(5) \times SO(2)_H$ (\emph{i.e.} K-type decomposition). There exists a unique lowest energy state in the Hilbert space of the minimal unitary representation of $SO(5,2)$ which is an $SO(5)$ singlet, namely
\begin{equation}
\ket{\Omega}
= | \psi_0^{(\alpha_g=1)} \rangle
= C_0 \, x \, e^{-x^2/2} \ket{0}
\end{equation}
with energy $E = \frac{3}{2}$. Note that $\ket{0}$ is the Fock vacuum of the $a$-type oscillators. The state $\ket{\Omega}$ is annihilated by all the operators $\widetilde{B}_1,\dots,\widetilde{B}_5$ in $\mathfrak{C}^-$ subspace of $\mathfrak{so}(5,2)$ given in equation (\ref{SO(5,2)3gradingC-}):
\begin{equation}
\widetilde{B}_M \ket{\Omega} = 0
\qquad \qquad (M=1,\dots,5)
\end{equation}
Therefore the minrep of $SO(5,2)$ is a unitary lowest weight representation. All the other states of the particle basis of the minrep with higher energies can be obtained from this state by repeatedly acting on it with the operators $\widetilde{B}_1^\dag,\dots,\widetilde{B}_5^\dag$ in $\mathfrak{C}^+$ subspace of $\mathfrak{so}(5,2)$ given in equation (\ref{SO(5,2)3gradingC+}):\footnote{The round brackets with a subscript $o$  in $B_{(M_1}^\dag \dots B_{M_n)_o}^\dag$ denote the symmetric traceless product of the operators.}
\begin{equation}
\ket{\Omega}
\quad , \quad
\widetilde{B}_{M_1}^\dag \ket{\Omega}
\quad , \quad
\widetilde{B}_{(M_1}^\dag \widetilde{B}_{M_2)_o}^\dag \ket{\Omega}
\quad , \quad
\widetilde{B}_{(M_1}^\dag \widetilde{B}_{M_2}^\dag \widetilde{B}_{M_3)_o}^\dag \ket{\Omega}
\quad , \quad
\dots\dots
\end{equation}

All the states of any given energy level form a single irrep of $SO(5) \approx U\!Sp(4)$. In Table \ref{Table:ScalarSingleton}, we give the decomposition of the minrep of $SO(5,2)$ with respect to its maximal compact subgroup by listing the  dimension of $SO(5) \approx U\!Sp(4)$ irrep and its $U\!Sp(4)$ Dynkin labels as well as their energies. The minrep of $SO(5,2)$ is the scalar singleton representation of $SO(5,2)$ in $AdS_6$ just like the Dirac scalar singleton of $SO(3,2)$ in $AdS_4$. It corresponds to a massless conformal scalar field  in five-dimensional Minkowski spacetime which can be identified with the boundary of $AdS_6$.  
%


\begin{small}
\begin{longtable}[c]{|c|c|c|c|}
\kill

\caption[The minimal unitary representation of $SO(5,2)$]
{K-type decomposition of the minimal unitary representation of $SO(5,2)$ with the lowest weight vector $\ket{\Omega} = C_0 \, x \, e^{-x^2/2} \ket{0}$. The $AdS$ energy, dimension of $SO(5) \approx U\!Sp(4)$ irrep and its $U\!Sp(4)$ Dynkin labels at each level are given.
\label{Table:ScalarSingleton}} \\
\hline
 & & & \\
States & $AdS_6$ & Dim of & Dynkin Labels \\
 & Energy & $SO(5) \approx U\!Sp(4)$ & of $U\!Sp(4)$ \\
 & $E$ & & \\
 & & & \\
\hline
 & & & \\
\endfirsthead
\caption[]{(continued)} \\
\hline
 & & & \\
States & $AdS$ & Dim of & Dynkin Labels \\
 & Energy & $SO(5) = U\!Sp(4)$ & of $U\!Sp(4)$ \\
 & $E$ & & \\
 & & & \\
\hline
\endhead
 & & & \\
\hline
\endlastfoot

$\ket{\Omega}$ &
$\frac{3}{2}$ &
1 &
$\left( 0 , 0 \right)$
\\[8pt]

\hline
 & & & \\

$\widetilde{B}_{M_1}^\dag \ket{\Omega}$ &
$\frac{5}{2}$ &
5 &
$\left( 0 , 1 \right)$
\\[8pt]

\hline
 & & & \\

$\widetilde{B}_{(M_1}^\dag \widetilde{B}_{M_2)_o}^\dag \ket{\Omega}$ &
$\frac{7}{2}$ &
14 &
$\left( 0 , 2 \right)$
\\[8pt]

\hline
 & & & \\

$\widetilde{B}_{(M_1}^\dag \widetilde{B}_{M_2}^\dag
   \widetilde{B}_{M_3)_o}^\dag \ket{\Omega}$ &
$\frac{9}{2}$ &
30 &
$\left( 0 , 3 \right)$
\\[8pt]

\hline
 & & & \\

\vdots & \vdots & \vdots & \vdots \\ [8pt]

\hline
 & & & \\
 
$\widetilde{B}_{(M_1}^\dag \dots \widetilde{B}_{M_n)_o}^\dag
   \ket{\Omega}$ &
$n + \frac{3}{2}$ &
$\frac{(n+1)(n+2)(2n+3)}{6}$ &
$\left( 0 , n \right)$
\\[8pt]

\hline
 & & & \\
 
\vdots & \vdots & \vdots & \vdots \\[8pt]

\end{longtable}
\end{small}

Note that the symmetric product $\widetilde{B}_{(M_1}^\dag \dots \widetilde{B}_{M_n)}^\dag$of the operators $B_M^\dag$ is automatically  traceless  due to the condition $\widetilde{B}_M^\dag \widetilde{B}_M^\dag = 0$ given in equation (\ref{MinrepC+Constraint}). When acting on the lowest weight state $\ket{\Omega}$, which is a singlet of $SO(5)$, it produces a state represented by an $SO(5)$ Young tableaux with one row of $n$ boxes.

\be
\underbrace{\begin{picture}(100,40)(01.8,-18)
\put(0,5){\line(0,1){10}}
\put(00,5){\line(1,0){100}}
\put(00,15){\line(1,0){100}}
\put(10,5){\line(0,1){10}}
\put(20,5){\line(0,1){10}}
\put(30,5){\line(0,1){10}}
\put(40,5){\line(0,1){10}}
\put(70,5){\line(0,1){10}}
\put(80,5){\line(0,1){10}}
\put(90,5){\line(0,1){10}}
\put(100,5){\line(0,1){10}}
\put(55,10){\makebox(0,0){$\cdots$}}
\end{picture}}_{\mbox{$n$ boxes}}
\ee


\section{Deformation of the minimal unitary representation of $SO(5,2)$} 
\label{sec:deformedSO(5,2)}

The little group of massless particles in five dimensions is $SU(2)$ and can be identified with the subgroup $SU(2)_L$ of $SO(5,2)$, considered as the five-dimensional conformal group. In this section we shall study possible ``deformations'' of the minrep of $SO(5,2)$, generated by adding a ``spin term'' $S_i$ to the ``orbital'' generators $L_i$ of $SU(2)_L$. We shall first introduce the ``spin'' generators as bilinears of two fermionic oscillators. This option is the natural one for extending the minrep of $SO(5,2)$ to that of the conformal supergroup $F(4)$ with the even subgroup $SO(5,2) \times SU(2)$.  

Consider the fermionic oscillators $\alpha_r$ ($r = 1,2$) and their hermitian conjugates $\alpha_r^\dag$ that satisfy the usual anti-commutation relations
\begin{equation}
\anticommute{\alpha_r}{\alpha_s^\dag}
= \delta_{rs}
\qquad \qquad \qquad
\anticommute{\alpha_r}{\alpha_s}
= \anticommute{\alpha_r^\dag}{\alpha_s^\dag} = 0
\label{FermionicOscillators}
\end{equation}
and define the 2-component spinor
\begin{equation}
\zeta = \left( \begin{matrix} \alpha_1 \\ \alpha_2 \end{matrix} \right) \,.
\end{equation}
We shall realize the generators $S_i$ of $\mathfrak{su}(2)_S$ as
\begin{equation}
S_i = \frac{1}{2} \, \zeta^\dag \, \sigma_i \, \zeta
\end{equation}
where $\sigma_i$ are the Pauli matrices. The quadratic Casimir of $\mathfrak{su}(2)_S$, denoted as $\mathcal{S}^2$, is given by
\begin{equation}
\mathcal{C}_2 \left[ \mathfrak{su}(2)_S \right]
= \mathcal{S}^2
= {S_1}^2 + {S_2}^2 + {S_3}^2 \,.
\end{equation}
Then the ``orbital'' generators $L_i$ of $\mathfrak{su}(2)_L$ get extended to the ``total angular momentum'' generators $J_i$ by adding the spin terms:
\begin{equation}
J_i = L_i + S_i
\label{SU(2)J}
\end{equation}
The quadratic Casimir of $\mathfrak{su}(2)_J$, denoted as $\mathcal{J}^2$, is given by
\begin{equation}
\mathcal{C}_2 \left[ \mathfrak{su}(2)_J \right]
= \mathcal{J}^2
= {J_1}^2 + {J_2}^2 + {J_3}^2
= \mathcal{L}^2 + \mathcal{S}^2 + 2 \, \mathcal{L} \cdot \mathcal{S}
\end{equation}
where $\mathcal{L} \cdot \mathcal{S} = L_1 S_1 + L_2 S_2 + L_3 S_3$.

This introduction of fermionic contributions does not affect the generators $M_{\pm,0}$ and $\Delta$ in grade 0 subspace, $U_i$ and $U_i^\dag$ in grade $-1$ subspace, and $K_-$ in grade $-2$ subspace of $\mathfrak{so}(5,2)$, defined in section {\ref{sec:minrepSO(5,2)}}. To preserve Jacobi identities, one finds that  the quadratic Casimir $\mathcal{L}^2$ appearing in grade $+2$ generator $K_+$ must be replaced by $\left( 2 \, \mathcal{J}^2 - \mathcal{L}^2 + \frac{2}{3} \, \mathcal{S}^2\right)$:
\begin{equation}
K_+ = \frac{1}{2} p^2
      + \frac{1}{2 \, x^2}
        \left(
         2 \, \mathcal{J}^2 - \mathcal{L}^2 + \frac{2}{3} \, \mathcal{S}^2
        \right)
\label{K+}
\end{equation}
Therefore the ``coupling constant'' of the isotonic (singular) oscillator becomes
\begin{equation}
\mathcal{G}
= \frac{1}{2}
  \left(
   2 \, \mathcal{J}^2 - \mathcal{L}^2 + \frac{2}{3} \mathcal{S}^2
  \right)
\end{equation}
and the  grade $+1$ generators, also changing according to equation (\ref{SO(5,2)grade+1}), become
\begin{equation}
\begin{split}
W_i
&= p \, a_i 
  - \frac{i}{x}
    \left[
     a_i + i \, \epsilon_{ijk} \left( L_j + 2 \, S_j \right) a_k
    \right]
\\
W_i^\dag
&= p \, a_i^\dag
   - \frac{i}{x}
     \left[
      a_i^\dag + i \, \epsilon_{ijk} \left( L_j + 2 \, S_j \right) a_k^\dag
     \right] \,.
\end{split}
\label{Grade+1Bosonic}
\end{equation}

All the commutation relations of $SO(5,2)$ given in the previous sections are still valid with the above replacements. The quadratic Casimir of the resulting realization of $\mathfrak{so}(5,2)$ with the fermionic contributions can be evaluated easily, and one finds 
\begin{equation}
\begin{split}
\mathcal{C}_2 \left[ \mathfrak{so}(5,2) \right]
&= \mathcal{C}_2 \left[ \mathfrak{su}(2)_J \right]
   + 2 \, \mathcal{C}_2 \left[ \mathfrak{su}(1,1)_M \right]
   - \frac{1}{2} \, \mathcal{C}_2 \left[ \mathfrak{su}(1,1)_K \right]
   + \frac{i}{2} \left[ UW \right]
\\
&=  \frac{7}{3} \mathcal{S}^2 - \frac{21}{4}
\end{split}
\end{equation}
where we have used 
\begin{equation}
\begin{split}
\mathcal{C}_2 \left[ \mathfrak{su}(1,1)_K \right]
&= \Delta^2 - 2 \left( K_+ K_- + K_- K_+ \right)
= \mathcal{K}^2
= - \left(
     2 \, \mathcal{J}^2 - \mathcal{L}^2 + \frac{2}{3} \mathcal{S}^2
    \right)
    + \frac{3}{4}
\\
\left[ U W \right]
&= U_i W_i^\dag + W_i^\dag U_i - U_i^\dag W_i - W_i U_i^\dag
= - 4 i \, \mathcal{K}^2 + 12 i
\\
&\qquad \qquad \qquad \qquad \qquad \qquad \quad \quad \;\;
= 4 i \left(
       2 \, \mathcal{J}^2 - \mathcal{L}^2 + \frac{2}{3} \mathcal{S}^2
      \right) + 9 i
\end{split}
\end{equation}
Thus the Casimir of the minrep, extended by fermionic contributions, depends only on the Casimir of $SU(2)_S$ generated by fermionic bilinears. The four-dimensional Fock space of the two fermionic oscillators decomposes as a doublet (spin $1/2$) and two singlets under the action of $SU(2)_S$. The singlets are the states
\begin{equation}
\ket{0}_F
\qquad  , \qquad
\alpha_1^\dag \, \alpha_2^\dag \ket{0}_F
\end{equation}
where $\ket{0}_F$ is the fermionic Fock vacuum. The states $\alpha_r^\dag \ket{0}_F$ ($r=1,2$) also form a doublet.
As a consequence, the tensor product space of the Fock spaces of the bosonic oscillators $a_i^\dag$ and the fermionic oscillators $\alpha_r^\dag$ with the state space of the singular oscillator deformed by the fermionic oscillators decomposes as two copies of the minrep and one copy of the deformed minrep as will be made evident in the next section. Here we should stress that increasing the number of fermionic oscillators so as to generate higher representations of $SU(2)_S$ leads to a failure of the Jacobi identities of $SO(5,2)$. As a consequence we find a single irreducible distinct deformation of the minrep of $SO(5,2)$, as in the case of $SO(3,2)$. This result is expected in the light of the results of \cite{Angelopoulos:1997ij}, where it was shown that there exist only two conformally massless fields in five dimensions. The 5-dimensional Poincar\'{e} mass operator vanishes identically
\begin{equation}
\mathcal{M}^2
= \eta_{\mu\nu} \mathcal{P}^\mu \mathcal{P}^\nu
= 0
\end{equation}
for the deformation of the minrep just as it did for the ``undeformed'' minrep. Thus we shall refer to the minrep as the scalar singleton and its deformation as the spinor singleton. They correspond to the conformally massless scalar and spinor fields in five dimensions.

Masslessness property of the momentum generators has a counterpart for the special conformal generators $\mathcal{K}^\mu$:
\begin{equation}
\eta_{\mu\nu} \mathcal{K}^\mu\mathcal{K}^\nu
= 0
\end{equation}
in both ``undeformed'' and ``deformed'' minrep.

The corresponding constraints in the compact 3-grading of $SO(5,2)$ with respect to the subgroup $SO(5) \times SO(2) \approx U\!Sp(4) \times U(1)$  are 
\begin{equation}
\widetilde{B}_M^\dag \widetilde{B}_M^\dag
= 0
\qquad \qquad
\Omega_{IK} \Omega_{JL} \overline{\mathcal{B}}_{IJ} \overline{\mathcal{B}}_{KL}
= 0
\end{equation}
and
\begin{equation}
\widetilde{B}_M \widetilde{B}_M
= 0
\qquad \qquad
\Omega_{IK} \Omega_{JL} \mathcal{B}_{IJ} \mathcal{B}_{KL}
= 0
\end{equation}
that are valid for both the minrep and its spinorial deformation.


\section{$K$-type decomposition of the deformed minimal unitary representation of $SO(5,2)$}
\label{sec:deformedminrep}

In four- and six-dimensional conformal algebras, the minrep of the conformal group corresponds to a massless scalar conformal field in the respective dimension, and one can obtain all the irreducible representations that correspond to other massless conformal fields  as ``deformations'' of the minrep \cite{Fernando:2009fq,Fernando:2010dp}.
In four dimensions one finds a continuous infinity of massless conformal fields labelled by helicity, and in six dimensions one finds a discrete infinity of massless conformal fields labelled by the spin $t$ of an $SU(2)_T$ subgroup of the little group $SO(4)$ of massless particles.

The five-dimensional conformal group $SO(5,2)$, on the other hand, admits \emph{only two} singleton representations corresponding to five-dimensional massless scalar and spinorial conformal fields. The minrep of $SO(5,2)$  corresponds to the scalar singleton representation (given in section \ref{sec:undeformedminrep}), and the ``deformation'' of the minrep labeled by the spin $s=1/2$ of the subgroup $SU(2)_S$ corresponds to the spinorial singleton representation. In this section we shall give the K-type decomposition of the deformed minrep. 

Consider the Fock space of the $a$-type bosonic oscillators and the $\alpha$-type fermionic oscillators introduced in previous sections, whose vacuum state $\ket{0}$ is annihilated by all $a_i$ and $\alpha_r$:
\begin{equation}
a_i \ket{0} = 0 \quad (i=1,2,3)
\qquad \qquad \qquad
\alpha_r \ket{0} = 0 \quad (r=1,2)
\end{equation}
A ``particle basis'' in this Fock space is provided by the states of the form
\begin{equation}
\ket{n_1,n_2,n_3 \,;\, \tilde{n}_1,\tilde{n}_2}
= \prod_i \prod_r \frac{1}{\sqrt{n_{i} !}} \, ( a_i^\dag )^{n_{i}} ( \alpha_r^\dag )^{\tilde{n}_r} \ket{0}
\end{equation}
where $n_{i}$ are non-negative integers, and $\tilde{n}_r$ can be either 0 or 1.
The Hilbert space of the deformed minimal unitary representation of $SO(5,2)$ is spanned by tensor product states of the form:
\begin{equation}
\ket{\psi_n^{(\alpha_g)}  \,;\, n_{1} , n_{2} , n_{3} \,;\, \tilde{n}_1,\tilde{n}_2}
\label{tensorstates}
\end{equation}

Since the ``spin'' $SU(2)_S$ is realized in terms of two fermionic oscillators, in the Hilbert space spanned by states of the form (\ref{tensorstates}), there are two states that are singlets of $U\!Sp(4)$ with a definite eigenvalue $E$ of $H$ and are annihilated by all grade $-1$ generators $\widetilde{B}_M$, namely, the state
\begin{equation}
\ket{\psi_0^{(1)}  \,;\, 0 , 0 , 0 \,;\, 0,0} 
= | \Phi_{0,0} \rangle
= x \, e^{-x^2/2} \ket{0}
\label{SO(5,2)LWV0_1}
\end{equation}
and the state 
\begin{equation}
\ket{\psi_0^{(1)}  \,;\, 0 , 0 , 0 \,;\, 1,1} 
= | \widetilde{\Phi}_{0,0} \rangle
= x \, e^{-x^2/2} \alpha_1^\dagger \alpha_2^\dagger \ket{0}
\label{SO(5,2)LWV0_2}
\end{equation}
They are the lowest weight vectors of two copies of the minrep of $SO(5,2)$ with different internal quantum numbers as will become manifest when we study the extension to the superalgebra $\mathfrak{f}(4)$.

In addition there exist four states that transform in the spinor representation of $U\!Sp(4)$ with a definite eigenvalue $E$ and are annihilated by all the grade $-1$ generators. They are:
\begin{equation}
\begin{split}
| \Psi_{+\frac{1}{2},0} \rangle
&= x^2 \, e^{-x^2/2} \, \alpha_1^\dag \ket{0}
   + \frac{1}{\sqrt{2}} \,
     x \, e^{-x^2/2}
     \left( \sigma_i^* \right)_{1s} a_i^\dag \alpha_s^\dag
     \ket{0}
\\
| \Psi_{-\frac{1}{2},0} \rangle
&= x^2 \, e^{-x^2/2} \, \alpha_2^\dag \ket{0}
   + \frac{1}{\sqrt{2}} \,
     x \, e^{-x^2/2}
     \left( \sigma_i^* \right)_{2s} a_i^\dag \alpha_s^\dag
     \ket{0}
\\
| \Psi_{0,+\frac{1}{2}} \rangle
&= x^2 \, e^{-x^2/2} \, \alpha_1^\dag \ket{0}
   - \frac{1}{\sqrt{2}}
     x \, e^{-x^2/2}
     \left( \sigma_i^* \right)_{1s} a_i^\dag \alpha_s^\dag
     \ket{0}
\\
| \Psi_{0,-\frac{1}{2}} \rangle
&= x^2 \, e^{-x^2/2} \, \alpha_2^\dag \ket{0}
   - \frac{1}{\sqrt{2}}
     x \, e^{-x^2/2}
     \left( \sigma_i^* \right)_{2s} a_i^\dag \alpha_s^\dag
     \ket{0}
\end{split}
\label{SO(5,2)LWV1/2}
\end{equation}
where the subscripts $0,\pm\frac{1}{2}$ in $\Phi$, $\widetilde{\Phi}$ and $\Psi$ refer to the respective eigenvalues of $A_3$ and $ \mathring{A}_3$ generators  of $SU(2)_A$ and $SU(2)_{\mathring{A}}$ subgroups of the rotation group $SO(4)$ in five dimensions as defined in equation (\ref{LeftRightSU(2)}). These  four states, which we shall denote as
\begin{equation}
\ket{\Omega_I}
= \left\{ | \Psi_{+\frac{1}{2},0} \rangle \,,\, | \Psi_{-\frac{1}{2},0} \rangle \,,\, | \Psi_{0,+\frac{1}{2}} \rangle \,,\, | \Psi_{0,-\frac{1}{2}} \rangle \right\} \,,
\end{equation}
form and irrep of  its compact subgroup $U\!Sp(4)$ with the lowest eigenvalue of $AdS_6$ energy. The corresponding unitary representation of $SO(5,2)$ is the unique deformation of the minrep.

The states of the ``particle basis'' of the deformed minrep are obtained by repeatedly acting on $\ket{\Omega_I}$ with grade $+1$ operators in the $\mathfrak{C}^+$ subspace of $SO(5,2)$.
\begin{equation}
\widetilde{B}_{M_1}^\dag \ket{\Omega_I}
\quad , \quad
\widetilde{B}_{(M_1}^\dag \widetilde{B}_{M_2)_o}^\dag \ket{\Omega_I}
\quad , \quad
\widetilde{B}_{(M_1}^\dag \widetilde{B}_{M_2}^\dag \widetilde{B}_{M_3)_o}^\dag \ket{\Omega_I}
\quad , \quad
\dots\dots
\end{equation}

In Table \ref{Table:SpinorSingleton}, we give the ``deformed'' minrep of $SO(5,2)$, with the corresponding $AdS$ energy, $SO(5) \approx U\!Sp(4)$ irreps and $U\!Sp(4)$ Dynkin labels. Clearly, this deformed minrep of $SO(5,2)$ corresponds to the spinor singleton representation of $SO(5,2)$, similar to the Dirac spinor singleton of $SO(3,2)$. In the context of the five-dimensional conformal group, it corresponds to a massless spinor field.

\begin{small}
\begin{longtable}[c]{|c|c|c|c|}
\kill

\caption[The deformed minimal unitary representation of $SO(5,2)$]
{K-type decomposition of the  deformed minimal unitary representation of $SO(5,2)$ obtained from the lowest energy irrep  $\ket{\Omega_I}$ of $USp(4)$. The $AdS_6$ energy, dimension of $SO(5) \approx U\!Sp(4)$ irrep and $U\!Sp(4)$ Dynkin labels of each level are given.
\label{Table:SpinorSingleton}} \\
\hline
 & & & \\
States & $AdS_6$ & Dim of & Dynkin Labels \\
 & Energy & $SO(5) \approx U\!Sp(4)$ & of $U\!Sp(4)$ \\
 & $E$ & & \\
 & & & \\
\hline
 & & & \\
\endfirsthead
\caption[]{(continued)} \\
\hline
 & & & \\
States & $AdS$ & Dim of & Dynkin Labels \\
 & Energy & $SO(5) \approx U\!Sp(4)$ & of $U\!Sp(4)$ \\
 & $E$ & & \\
 & & & \\
\hline
\endhead
 & & & \\
\hline
\endfoot
 & & & \\
\hline
\endlastfoot

$\ket{\Omega_I}$ &
2 &
4 &
$\left( 1 , 0 \right)$
\\[8pt]

\hline
 & & & \\

$\widetilde{B}_{M_1}^\dag \ket{\Omega_I}$ &
3 &
16 &
$\left( 1 , 1 \right)$
\\[8pt]

\hline
 & & & \\

$\widetilde{B}_{(M_1}^\dag \widetilde{B}_{M_2)_o}^\dag \ket{\Omega_I}$ &
4 &
40 &
$\left( 1 , 2 \right)$
\\[8pt]

\hline
 & & & \\

$\widetilde{B}_{(M_1}^\dag \widetilde{B}_{M_2}^\dag \widetilde{B}_{M_3)_o}^\dag \ket{\Omega_I}$ &
5 &
80 &
$\left( 1 , 3 \right)$
\\[8pt]

\hline
 & & & \\

\vdots & \vdots & \vdots & \vdots \\ [8pt]

\hline
 & & & \\
 
$\widetilde{B}_{(M_1}^\dag \dots \widetilde{B}_{M_n)_o}^\dag \ket{\Omega_I}$ &
$n + 2$ &
$\frac{2(n+1)(n+2)(n+3)}{3}$ &
$\left( 1 , n \right)$
\\[8pt]

\hline
 & & & \\
 
\vdots & \vdots & \vdots & \vdots \\

\end{longtable}
\end{small}


\section{Minimal unitary representation of the exceptional Lie superalgebra  $F(4)$} 
\label{sec:F(4)minrep}

The superalgebra $\mathfrak{f}(4)$ with the even subalgebra $\mathfrak{so}(5,2) \oplus \mathfrak{su}(2)$ is the unique simple conformal superalgebra in five dimensions. To construct the minimal unitary irreducible representation of  $\mathfrak{f}(4)$ via the quasiconformal method, we start from its 5-graded decomposition with respect to its subsuperalgebra $d(2,1;2) \oplus \mathfrak{so}(1,1)$:
\begin{equation}
\begin{split}
\mathfrak{f}(4)
&= \mathfrak{g}^{(-2)} \oplus
   \mathfrak{g}^{(-1)} \oplus
   \mathfrak{g}^{(0)} \oplus
   \mathfrak{g}^{(+1)} \oplus
   \mathfrak{g}^{(+2)}
\\
&= 1_B \oplus
   \left( 6_B \oplus 4_F \right) \oplus
   \left[ d(2,1;2) \oplus \Delta \right] \oplus
   \left( 6_B \oplus 4_F \right) \oplus
   1_B
\end{split}
\end{equation}
The subsuperalgebra $d(2,1;2)$ belongs to the continuous family of 17-dimensional Lie superalgebras $d(2,1;\alpha)$, and admits a 10-dimensional linear representation with 6 bosonic and 4 fermionic degrees of freedom. Grade $\pm 1$ generators of $\mathfrak{f}(4)$ transform in this 10-dimensional representation as indicated above. In our case the relevant real form of $d(2,1;2)$ has the even subalgebra $\mathfrak{su}(2) \oplus \mathfrak{su}(2) \oplus \mathfrak{su}(1,1)$. 

In a previous section, we constructed the deformation of the minimal unitary representation of $SO(5,2)$ by using $\alpha$-type fermionic oscillators. Now we shall realize the $SU(2)$ subgroup outside $SO(5,2)$, which acts as the $R$-symmetry group, entirely in terms of the same fermionic oscillators as follows:
\begin{equation}
T_1
= \frac{1}{2} \left( \alpha_1^\dag \alpha_2^\dag + \alpha_2 \alpha_1 \right)
\qquad
T_2
= - \frac{i}{2} \left( \alpha_1^\dag \alpha_2^\dag - \alpha_2 \alpha_1 \right)
\qquad
T_3
= \frac{1}{2} \left( \alpha_1^\dag \alpha_1 - \alpha_2 \alpha_2^\dag \right)
\label{SU(2)T}
\end{equation}
They satisfy the commutation relations:
\begin{equation}
\commute{T_i}{T_j} = i \, \epsilon_{ijk} \, T_k
\end{equation}
The raising and lowering operators of this subalgebra will be denoted as  $T_\pm = T_1 \pm i \, T_2$. We shall denote this subalgebra as $\mathfrak{su}(2)_T$ and its quadratic Casimir as $\mathcal{T}^2$:
\begin{equation}
\mathcal{C}_2 \left[ \mathfrak{su}(2)_T \right]
= \mathcal{T}^2
= {T_1}^2 + {T_2}^2 + {T_3}^2
\end{equation}
This quadratic Casimir is related to that of $\mathfrak{su}(2)_S$ as
\begin{equation}
\mathcal{T}^2 = \frac{3}{4} - \mathcal{S}^2 \,.
\end{equation}

We shall use the generator $\Delta = \frac{1}{2} \left( x p + p x \right)$, first given in equation (\ref{delta}), as the operator that defines the 5-grading, and identify the 6 even generators in grade $-1$ subspace of $\mathfrak{f}(4)$ with the generators $U_i = x \, a_i$ and $U_i^\dag = x \, a_i^\dag$  of $\mathfrak{so}(5,2)$ given in equation (\ref{Grade-1Bosonic}).
Grade $-2$ generator $K_- = \frac{1}{2} x^2$, given in equation (\ref{K-}), will also remain unchanged in the extension to $\mathfrak{f}(4)$.

The four supersymmetry operators in grade $-1$ subspace will be defined as follows:
\begin{equation}
Q_r = x \, \alpha_r
\qquad \qquad \qquad
Q_r^\dag = x \, \alpha_r^\dag
\label{Grade-1Fermionic}
\end{equation}
Under anti-commutation they form a super-Heisenberg algebra that closes into the grade $-2$ generator $K_-$.
\begin{equation}
\anticommute{Q_r}{Q_s^\dag} = 2 \, \delta_{rs} \, K_-
\qquad \qquad
\anticommute{Q_r}{Q_s} = \anticommute{Q_r^\dag}{Q_s^\dag} = 0
\end{equation}
Clearly $( Q_1 , Q_2 )$ and $( Q_2^\dag , Q_1^\dag )$ form doublets under $SU(2)_J$, while $( Q_1 , Q_2^\dag )$ and $( Q_2 , Q_1^\dag )$ form doublets under the $R$-symmetry group $SU(2)_T$.

Grade $+2$ generator $K_+$ also remains unchanged in the extension to $\mathfrak{f}(4)$, as given in equation (\ref{K+}), and the six even generators in grade $+1$ space are identified with the generators $W_i$ and $W_i^\dag$ of $\mathfrak{so}(5,2)$ given in equation (\ref{Grade+1Bosonic}).

Now the four supersymmetry operators in grade $+1$ subspace can be obtained by taking commutators between the generator $K_+$ in grade $+2$ subspace and the supersymmetry operators in grade $-1$ subspace:
\begin{equation}
R_r
= - i \commute{Q_r}{K_+}
\qquad \qquad \qquad
R_r^\dag
= - i \commute{Q_r^\dag}{K_+}
\label{SusyGrade+1}
\end{equation}
They have the following explicit form:
\begin{equation}
\begin{split}
R_r
&= p \, \alpha_r
   - \frac{i}{x}
     \left[
      \alpha_r
      + \left( \sigma_i \right)_{rs}
        \left( L_i + \frac{4}{3} S_i \right) \alpha_s
     \right]
\\
R_r^\dag
&= p \, \alpha_r^\dag
   - \frac{i}{x}
     \left[
      \alpha_r^\dag
      - \left( \sigma_i^* \right)_{rs}
        \left( L_i + \frac{4}{3} S_i \right) \alpha_s^\dag
     \right]
\end{split}
\end{equation}

Commutators between these supersymmetry operators in grade $+1$ subspace and the generator $K_-$ in grade $-2$ subspace produce the respective supersymmetry operators in grade $-1$ subspace:
\begin{equation}
\commute{R_r}{K_-} = - i \, Q_r
\qquad \qquad
\commute{R_r^\dag}{K_-} = - i \, Q_r^\dag
\end{equation}
Under anti-commutation, these four supersymmetry operators in grade $+1$ subspace generate  a super-Heisenberg algebra that closes into the grade $+2$ generator $K_+$.
\begin{equation}
\anticommute{R_r}{R_s^\dag} = 2 \, \delta_{rs} \, K_+
\qquad \qquad
\anticommute{R_r}{R_s} = \anticommute{R_r^\dag}{R_s^\dag} = 0
\end{equation}
$( R_1 , R_2 )$ and $( R_2^\dag , R_1^\dag )$ form doublets under $SU(2)_J$, while $( R_1 , R_2^\dag )$ and $( R_2 , R_1^\dag )$ form doublets under the $R$-symmetry group $SU(2)_T$.

The anti-commutators between the supersymmetry operators in grade $-1$ subspace and those in grade $+1$ subspace close into even generators in grade 0 subspace:
\begin{equation}
\begin{aligned}
\anticommute{Q_r}{R_s}
&= - 3 i \, \epsilon_{rs} \, T_-
\\
\anticommute{Q_r^\dag}{R_s^\dag}
&= 3 i \, \epsilon_{rs} \, T_+
\end{aligned}
\qquad \qquad
\begin{aligned}
\anticommute{Q_r}{R_s^\dag}
&= \delta_{rs} \left( \Delta - 3 i \, T_3 \right)
   + i \left( \sigma_i \right)_{rs} \, J_i
\\
\anticommute{Q_r^\dag}{R_s}
&= \delta_{rs} \left( \Delta + 3 i \, T_3 \right)
   - i \left( \sigma_i^* \right)_{rs} \, J_i
\end{aligned}
\end{equation}

With respect to the compact generators $J_3$ and $T_3$, these eight supersymmetry operators in grade $\pm 1$ subspaces have the charges given in Table \ref{Table:Grade1SusyCharges}, which show their transformation properties under $SU(2)_J \times SU(2)_T$.

The remaining eight supersymmetry operators of $\mathfrak{f}(4)$ reside in grade 0 subspace, and they can be obtained by taking commutators between supersymmetry operators in grade $-1$ (grade $+1$) subspace and even generators in grade $+1$ (grade $-1$) subspace. These eight supersymmetry operators are bilinears of the $a$-type bosonic oscillators and the $\alpha$-type fermionic oscillators and are given as follows:
\begin{equation}
\begin{aligned}
\Sigma_r
&= \left( \sigma_i \right)_{rs} a_i \, \alpha_s \\
\Sigma_r^\dag
&= \left( \sigma_i^* \right)_{rs} a_i^\dag \, \alpha_s^\dag
\end{aligned}
\qquad \qquad
\begin{aligned}
\Pi_r
&= \left( \sigma_i \right)_{rs} a_i^\dag \, \alpha_s
\\
\Pi_r^\dag
&= \left( \sigma_i^* \right)_{rs} a_i \, \alpha_s^\dag
\end{aligned}
\end{equation}

With respect to the compact generators $J_3$ and $T_3$ and $M_0$, these eight supersymmetry operators in grade 0 subspace have the charges given in Table \ref{Table:Grade0SusyCharges}, which show their transformation properties under $SU(2)_J \times SU(2)_T \times SU(1,1)_M$.

\begin{small}
\begin{longtable}[c]{|c||c|c|c|c|c|c|c|c|}
\kill

\caption[$J_3$ and $T_3$ charges of grade $\pm1$ supersymmetry operators]
{$J_3$ and $T_3$ charges of grade $\pm1$ supersymmetry operators
\label{Table:Grade1SusyCharges}} \\
\hline
& & & & & & & & \\
& $Q_1$ & $Q_2$ & $Q_1^\dag$ & $Q_2^\dag$ & $R_1$ & $R_2$ & $R_1^\dag$ & $R_2^\dag$ \\
& & & & & & & & \\
\hline
& & & & & & & & \\
\endfirsthead
\caption[]{(continued)} \\
\hline
& & & & & & & & \\
& $Q_1$ & $Q_2$ & $Q_1^\dag$ & $Q_2^\dag$ & $R_1$ & $R_2$ & $R_1^\dag$ & $R_2^\dag$ \\
& & & & & & & & \\
\hline
& & & & & & & & \\
\endhead
& & & & & & & & \\
\hline
\endfoot
& & & & & & & & \\
\hline
\endlastfoot

$J_3$ & $-\frac{1}{2}$ & $+\frac{1}{2}$ & $+\frac{1}{2}$ & $-\frac{1}{2}$ & $-\frac{1}{2}$ & $+\frac{1}{2}$ & $+\frac{1}{2}$ & $-\frac{1}{2}$
\\[8pt]

$T_3$ & $-\frac{1}{2}$ & $-\frac{1}{2}$ & $+\frac{1}{2}$ & $+\frac{1}{2}$ & $-\frac{1}{2}$ & $-\frac{1}{2}$ & $+\frac{1}{2}$ & $+\frac{1}{2}$
\\[8pt]


\end{longtable}
\end{small}

\begin{small}
\begin{longtable}[c]{|c||c|c|c|c|c|c|c|c|}
\kill

\caption[$J_3$, $T_3$ and $M_0$ charges of grade $0$ supersymmetry operators]
{$J_3$, $T_3$ and $M_0$ charges of grade $0$ supersymmetry operators
\label{Table:Grade0SusyCharges}} \\
\hline
& & & & & & & & \\
& $\Sigma_1$ & $\Sigma_2$ & $\Sigma_1^\dag$ & $\Sigma_2^\dag$ & $\Pi_1$ & $\Pi_2$ & $\Pi_1^\dag$ & $\Pi_2^\dag$ \\
& & & & & & & & \\
\hline
& & & & & & & & \\
\endfirsthead
\caption[]{(continued)} \\
\hline
& & & & & & & & \\
& $\Sigma_1$ & $\Sigma_2$ & $\Sigma_1^\dag$ & $\Sigma_2^\dag$ & $\Pi_1$ & $\Pi_2$ & $\Pi_1^\dag$ & $\Pi_2^\dag$ \\
& & & & & & & & \\
\hline
& & & & & & & & \\
\endhead
& & & & & & & & \\
\hline
\endfoot
& & & & & & & & \\
\hline
\endlastfoot

$J_3$ & $-\frac{1}{2}$ & $+\frac{1}{2}$ & $+\frac{1}{2}$ & $-\frac{1}{2}$ & $-\frac{1}{2}$ & $+\frac{1}{2}$ & $+\frac{1}{2}$ & $-\frac{1}{2}$
\\[8pt]

$T_3$ & $-\frac{1}{2}$ & $-\frac{1}{2}$ & $+\frac{1}{2}$ & $+\frac{1}{2}$ & $-\frac{1}{2}$ & $-\frac{1}{2}$ & $+\frac{1}{2}$ & $+\frac{1}{2}$
\\[8pt]

$M_0$ & $-\frac{1}{2}$ & $-\frac{1}{2}$ & $+\frac{1}{2}$ & $+\frac{1}{2}$ & $+\frac{1}{2}$ & $+\frac{1}{2}$ & $-\frac{1}{2}$ & $-\frac{1}{2}$
\\

\end{longtable}
\end{small}

The eight supersymmetry operators $\Sigma_r$, $\Sigma_r^\dag$, $\Pi_r$ and $\Pi_r^\dag$ in grade 0 subspace, along with the generators of $SU(2)_J$, $SU(1,1)_M$ and $SU(2)_T$, form the subsuperalgebra $d(2,1;2)$, which is in the grade 0 subspace of $\mathfrak{f}(4)$ in this 5-graded decomposition with respect to $\Delta$.

Finally, we give below some of the remaining (super-)commutators of $\mathfrak{f}(4)$ in this 5-grading:
\begin{subequations}
\begin{equation}
\begin{aligned}
\commute{U_i}{R_r}
&= i \left( \sigma_i \right)_{rs} \Sigma_s
\\
\commute{U_i}{R_r^\dag}
&= i \left( \sigma_i^* \right)_{rs} \Pi_s^\dag
\end{aligned}
\qquad \qquad \qquad
\begin{aligned}
\commute{Q_r}{W_i}
&= i \left( \sigma_i \right)_{rs} \Sigma_s
\\
\commute{Q_r^\dag}{W_i}
&= i \left( \sigma_i^* \right)_{rs} \Pi_s^\dag
\end{aligned}
\end{equation}
\begin{equation}
\begin{aligned}
\commute{U_i}{\Sigma_r}
&= 0
\\
\commute{U_i}{\Sigma_r^\dag}
&= \left( \sigma_i^* \right)_{rs} Q_s^\dag
\\
\commute{U_i}{\Pi_r}
&= \left( \sigma_i \right)_{rs} Q_s
\\
\commute{U_i}{\Pi_r^\dag}
&= 0
\end{aligned}
\qquad \qquad \qquad
\begin{aligned}
\commute{W_i}{\Sigma_r}
&= 0
\\
\commute{W_i}{\Sigma_r^\dag}
&= \left( \sigma_i^* \right)_{rs} R_s^\dag
\\
\commute{W_i}{\Pi_r}
&= \left( \sigma_i \right)_{rs} R_s
\\
\commute{W_i}{\Pi_r^\dag}
&= 0
\end{aligned}
\end{equation}
\begin{equation}
\begin{aligned}
\anticommute{Q_r}{\Sigma_s}
&= 0
\\\anticommute{Q_r}{\Sigma_s^\dag}
&= \left( \sigma_i \right)_{rs} U_i^\dag
\\
\anticommute{Q_r}{\Pi_s}
&= 0
\\\anticommute{Q_r}{\Pi_s^\dag}
&= \left( \sigma_i \right)_{rs} U_i
\end{aligned}
\qquad \qquad \qquad
\begin{aligned}
\anticommute{R_r}{\Sigma_s}
&= 0
\\\anticommute{R_r}{\Sigma_s^\dag}
&= \left( \sigma_i \right)_{rs} W_i^\dag
\\
\anticommute{R_r}{\Pi_s}
&= 0
\\\anticommute{R_r}{\Pi_s^\dag}
&= \left( \sigma_i \right)_{rs} W_i
\end{aligned}
\end{equation}
\end{subequations}


\section{Compact 3-grading of $\mathfrak{f}(4)$ with respect to the subsuperalgebra $\mathfrak{osp}(2|4) \oplus \mathfrak{u}(1)$}
\label{sec:F(4)c3G}

The Lie superalgebra $\mathfrak{f}(4)$ can be given a 3-graded decomposition with respect to its compact subsuperalgebra $\mathfrak{C}^0 = \mathfrak{osp}(2|4) \oplus \mathfrak{u}(1)_{\mathcal{H}}$
\begin{equation}
\mathfrak{f}(4)
= \mathfrak{C}^- \oplus \mathfrak{C}^0 \oplus \mathfrak{C}^+
\end{equation}
where
\begin{equation}
\begin{split}
\mathfrak{C}^-
&= \left( \widetilde{B}_1 , \dots , \widetilde{B}_5 \right) \oplus
   T_- \oplus
   \frac{1}{2} \left( Q_r + i \, R_r \right) \oplus
   \Sigma_r
\\
\mathfrak{C}^0
&= \mathcal{H} \oplus
   \widetilde{M}_{MN} \oplus
   \left[ \frac{1}{2} \left( K_+ + K_- \right) + M_0 + 3 \, T_3 \right]
\\
& \quad \oplus
   \frac{1}{2} \left( Q_r - i \, R_r \right) \oplus
   \frac{1}{2} \left( Q_r^\dag + i \, R_r^\dag \right) \oplus
   \Pi_r \oplus
   \Pi_r^\dag
\\
\mathfrak{C}^+
&= \left( \widetilde{B}_1^\dag , \dots , \widetilde{B}_5^\dag \right) \oplus
   T_+ \oplus
   \frac{1}{2} \left( Q_r^\dag - i \, R_r^\dag \right) \oplus
   \Sigma_r^\dag
\end{split}
\label{F(4)3grading}
\end{equation}
Note that $\widetilde{B}_M$, $\widetilde{B}_M^\dag$ and $\widetilde{M}_{MN}$ are the generators that formed $\mathfrak{so}(5,2)$.

The $\mathfrak{u}(1)$ generator $\mathcal{H}$ that defines the compact 3-grading of $\mathfrak{f}(4)$ is given by
\begin{equation}
\begin{split}
\mathcal{H}
&= \frac{1}{2} \left( K_+ + K_- \right)
    + M_0 + T_3
\\
&= \frac{1}{4} \left( x - i \, p \right) \left( x + i \, p \right)
   + \frac{1}{2 \, x^2} G
   + \frac{1}{2} \, a_i^\dag a_i
   + \frac{1}{2} \, \alpha_r^\dag \alpha_r
   + \frac{1}{2} \,.
\end{split}
\label{SuperHamiltonian}
\end{equation}

The subsuperalgebra $\mathfrak{osp}(2|4)$ in $\mathfrak{C}^0$ subspace has the even subalgebra of $\mathfrak{so}(2) \oplus \mathfrak{usp}(4)$. We shall denote this $\mathfrak{so}(2) \approx \mathfrak{u}(1)$ generator as
\begin{equation}
Z = \frac{1}{2} \left( K_+ + K_- \right) + M_0 + 3 \, T_3 \,.
\end{equation}

Grade 0 supersymmetry generators, \emph{i.e.} those that are in the subsuperalgebra $\mathfrak{osp}(2|4)$, shall be denoted as $\mathcal{R}_I$ and $\overline{\mathcal{R}}_I$ where 
\begin{equation}
\overline{\mathcal{R}}_I
= \mathcal{R}_J^\dag \Omega_{JI}
\end{equation}
Their explicit forms are given by
\begin{subequations}
\begin{equation}
\begin{split}
\mathcal{R}_1
&= \left[
    \frac{1}{2} \left( Q_1 - i \, R_1 \right)
    - \frac{1}{\sqrt{2}} \Pi_1
   \right]
\\
&= \frac{1}{2} \left( x - i \, p \right) \alpha_1
   - \frac{1}{2x}
     \left[
      \alpha_1
      + \left( \sigma_i \right)_{1s}
        \left( L_i + \frac{4}{3} S_i \right) \alpha_s
      \right]
   - \frac{1}{\sqrt{2}} \left( \sigma_i \right)_{1s} a_i^\dag \alpha_s
\\
\mathcal{R}_2
&= - \left[
      \frac{1}{2} \left( Q_2 - i \, R_2 \right)
      - \frac{1}{\sqrt{2}} \Pi_2
     \right]
\\
&= - \frac{1}{2} \left( x - i \, p \right) \alpha_2
   + \frac{1}{2x}
     \left[
      \alpha_2
      + \left( \sigma_i \right)_{2s}
        \left( L_i + \frac{4}{3} S_i \right) \alpha_s
      \right]
   + \frac{1}{\sqrt{2}} \left( \sigma_i \right)_{2s} a_i^\dag \alpha_s
\\
\mathcal{R}_3
&= i \left[
      \frac{1}{2} \left( Q_1 - i \, R_1 \right)
      + \frac{1}{\sqrt{2}} \Pi_1
     \right]
\\
&= \frac{i}{2} \left( x - i \, p \right) \alpha_1
   - \frac{i}{2x}
     \left[
      \alpha_1
      + \left( \sigma_i \right)_{1s}
        \left( L_i + \frac{4}{3} S_i \right) \alpha_s
      \right]
   + \frac{i}{\sqrt{2}} \left( \sigma_i \right)_{1s} a_i^\dag \alpha_s
\\
\mathcal{R}_4
&= - i \left[
        \frac{1}{2} \left( Q_2 - i \, R_2 \right)
        + \frac{1}{\sqrt{2}} \Pi_2
       \right]
\\
&= - \frac{i}{2} \left( x - i \, p \right) \alpha_2
   + \frac{i}{2x}
     \left[
      \alpha_2
      + \left( \sigma_i \right)_{2s}
        \left( L_i + \frac{4}{3} S_i \right) \alpha_s
      \right]
   - \frac{i}{\sqrt{2}} \left( \sigma_i \right)_{2s} a_i^\dag \alpha_s
\end{split}
\end{equation}
\begin{equation}
\begin{split}
\overline{\mathcal{R}}_1
&= \left[
    \frac{1}{2} \left( Q_2^\dag + i \, R_2^\dag \right)
    - \frac{1}{\sqrt{2}} \Pi_2^\dag
   \right]
\\
&= \frac{1}{2} \left( x + i \, p \right) \alpha_2^\dag
   + \frac{1}{2x}
     \left[
      \alpha_2^\dag
      - \left( \sigma_i^* \right)_{2s}
        \left( L_i + \frac{4}{3} S_i \right) \alpha_s^\dag
      \right]
   - \frac{1}{\sqrt{2}} \left( \sigma_i^* \right)_{2s} a_i \alpha_s^\dag
\\
\overline{\mathcal{R}}_2
&= \left[
    \frac{1}{2} \left( Q_1^\dag + i \, R_1^\dag \right)
    - \frac{1}{\sqrt{2}} \Pi_1^\dag
   \right]
\\
&= \frac{1}{2} \left( x + i \, p \right) \alpha_1^\dag
   + \frac{1}{2x}
     \left[
      \alpha_1^\dag
      - \left( \sigma_i^* \right)_{1s}
        \left( L_i + \frac{4}{3} S_i \right) \alpha_s^\dag
      \right]
   - \frac{1}{\sqrt{2}} \left( \sigma_i^* \right)_{1s} a_i \alpha_s^\dag
\\
\overline{\mathcal{R}}_3
&= - i \left[
        \frac{1}{2} \left( Q_2^\dag + i \, R_2^\dag \right)
        + \frac{1}{\sqrt{2}} \Pi_2^\dag
       \right]
\\
&= - \frac{i}{2} \left( x + i \, p \right) \alpha_2^\dag
   - \frac{i}{2x}
     \left[
      \alpha_2^\dag
      - \left( \sigma_i^* \right)_{2s}
        \left( L_i + \frac{4}{3} S_i \right) \alpha_s^\dag
      \right]
   - \frac{i}{\sqrt{2}} \left( \sigma_i^* \right)_{2s} a_i \alpha_s^\dag
\\
\overline{\mathcal{R}}_4
&= - i \left[
        \frac{1}{2} \left( Q_1^\dag + i \, R_1^\dag \right)
        + \frac{1}{\sqrt{2}} \Pi_1^\dag
       \right]
\\
&= - \frac{i}{2} \left( x + i \, p \right) \alpha_1^\dag
   - \frac{i}{2x}
     \left[
      \alpha_1^\dag
      - \left( \sigma_i^* \right)_{1s}
        \left( L_i + \frac{4}{3} S_i \right) \alpha_s^\dag
      \right]
   - \frac{i}{\sqrt{2}} \left( \sigma_i^* \right)_{1s} a_i \alpha_s^\dag
\end{split}
\end{equation}
\end{subequations}

The canonical commutation relations of the $\mathfrak{osp}(2|4)$ subsuperalgebra are as follows: 
\begin{equation}
\begin{split}
\commute{U_{IJ}}{U_{KL}}
&= \Omega_{JK} \, U_{IL} + \Omega_{IK} \, U_{JL}
   + \Omega_{JL} \, U_{IK} + \Omega_{IL} \, U_{JK}
\\
\anticommute{\mathcal{R}_I}{\mathcal{R}_J}
&= 0
\\
\anticommute{\overline{\mathcal{R}}_I}{\overline{\mathcal{R}}_J}
&= 0
\\
\anticommute{\mathcal{R}_I}{\overline{\mathcal{R}}_J}
&= \Omega_{IJ} \, Z - U_{IJ}
\\
\commute{\mathcal{Z}}{\mathcal{R}_I}
&= - \mathcal{R}_I
\\
\commute{\mathcal{Z}}{\overline{\mathcal{R}}_I}
&= + \overline{\mathcal{R}}_I
\\
\commute{U_{IJ}}{\mathcal{R}_K}
&= \Omega_{JK} \, \mathcal{R}_I + \Omega_{IK} \, \mathcal{R}_J
\\
\commute{U_{IJ}}{\overline{\mathcal{R}}_K}
&= \Omega_{JK} \, \overline{\mathcal{R}}_I
   + \Omega_{IK} \, \overline{\mathcal{R}}_J
\end{split}
\end{equation}

The supersymmetry operators that belong to grade $-1$ subspace and those that belong to grade $+1$ subspace shall be denoted by $\mathcal{Q}_I$ and $\overline{\mathcal{Q}}_I$, respectively, where 
\begin{equation}
\overline{\mathcal{Q}}_I
= \mathcal{Q}_J^\dag \Omega_{JI}
\end{equation}
which have the following explicit expressions:
\begin{subequations}
\begin{equation}
\begin{split}
\mathcal{Q}_1
&= \left[
    \frac{1}{2} \left( Q_1 + i \, R_1 \right)
    + \frac{1}{\sqrt{2}} \Sigma_1
   \right]
\\
&= \frac{1}{2} \left( x + i \, p \right) \alpha_1
   + \frac{1}{2x}
     \left[
      \alpha_1
      + \left( \sigma_i \right)_{1s}
        \left( L_i + \frac{4}{3} S_i \right) \alpha_s
      \right]
   + \frac{1}{\sqrt{2}} \left( \sigma_i \right)_{1s} a_i \alpha_s
\\
\mathcal{Q}_2
&= - \left[
      \frac{1}{2} \left( Q_2 + i \, R_2 \right)
      + \frac{1}{\sqrt{2}} \Sigma_2
     \right]
\\
&= - \frac{1}{2} \left( x + i \, p \right) \alpha_2
   - \frac{1}{2x}
     \left[
      \alpha_2
      + \left( \sigma_i \right)_{2s}
        \left( L_i + \frac{4}{3} S_i \right) \alpha_s
      \right]
   - \frac{1}{\sqrt{2}} \left( \sigma_i \right)_{2s} a_i \alpha_s
\\
\mathcal{Q}_3
&= - i \left[
        \frac{1}{2} \left( Q_1 + i \, R_1 \right)
        - \frac{1}{\sqrt{2}} \Sigma_1
       \right]
\\
&= - \frac{i}{2} \left( x + i \, p \right) \alpha_1
   - \frac{i}{2x}
     \left[
      \alpha_1
      + \left( \sigma_i \right)_{1s}
        \left( L_i + \frac{4}{3} S_i \right) \alpha_s
      \right]
   + \frac{i}{\sqrt{2}} \left( \sigma_i \right)_{1s} a_i \alpha_s
\\
\mathcal{Q}_4
&= i \left[
      \frac{1}{2} \left( Q_2 + i \, R_2 \right)
      - \frac{1}{\sqrt{2}} \Sigma_2
     \right]
\\
&= \frac{i}{2} \left( x + i \, p \right) \alpha_2
   + \frac{i}{2x}
     \left[
      \alpha_2
      + \left( \sigma_i \right)_{2s}
        \left( L_i + \frac{4}{3} S_i \right) \alpha_s
      \right]
   - \frac{i}{\sqrt{2}} \left( \sigma_i \right)_{2s} a_i \alpha_s
\end{split}
\end{equation}
\begin{equation}
\begin{split}
\overline{\mathcal{Q}}_1
&= \left[
    \frac{1}{2} \left( Q_2^\dag - i \, R_2^\dag \right)
    + \frac{1}{\sqrt{2}} \Sigma_2^\dag
   \right]
\\
&= \frac{1}{2} \left( x - i \, p \right) \alpha_2^\dag
   - \frac{1}{2x}
     \left[
      \alpha_2^\dag
      - \left( \sigma_i^* \right)_{2s}
        \left( L_i + \frac{4}{3} S_i \right) \alpha_s^\dag
      \right]
   + \frac{1}{\sqrt{2}} \left( \sigma_i^* \right)_{2s} a_i^\dag \alpha_s^\dag
\\
\overline{\mathcal{Q}}_2
&= \left[
    \frac{1}{2} \left( Q_1^\dag - i \, R_1^\dag \right)
    + \frac{1}{\sqrt{2}} \Sigma_1^\dag
   \right]
\\
&= \frac{1}{2} \left( x - i \, p \right) \alpha_1^\dag
   - \frac{1}{2x}
     \left[
      \alpha_1^\dag
      - \left( \sigma_i^* \right)_{1s}
        \left( L_i + \frac{4}{3} S_i \right) \alpha_s^\dag
      \right]
   + \frac{1}{\sqrt{2}} \left( \sigma_i^* \right)_{1s} a_i^\dag \alpha_s^\dag
\\
\overline{\mathcal{Q}}_3
&= i \left[
      \frac{1}{2} \left( Q_2^\dag - i \, R_2^\dag \right)
      - \frac{1}{\sqrt{2}} \Sigma_2^\dag
     \right]
\\
&= \frac{i}{2} \left( x - i \, p \right) \alpha_2^\dag
   - \frac{i}{2x}
     \left[
      \alpha_2^\dag
      - \left( \sigma_i^* \right)_{2s}
        \left( L_i + \frac{4}{3} S_i \right) \alpha_s^\dag
      \right]
   - \frac{i}{\sqrt{2}} \left( \sigma_i^* \right)_{2s} a_i^\dag \alpha_s^\dag
\\
\overline{\mathcal{Q}}_4
&= i \left[
      \frac{1}{2} \left( Q_1^\dag - i \, R_1^\dag \right)
      - \frac{1}{\sqrt{2}} \Sigma_1^\dag
     \right]
\\
&= \frac{i}{2} \left( x - i \, p \right) \alpha_1^\dag
   - \frac{i}{2x}
     \left[
      \alpha_1^\dag
      - \left( \sigma_i^* \right)_{1s}
        \left( L_i + \frac{4}{3} S_i \right) \alpha_s^\dag
      \right]
   - \frac{i}{\sqrt{2}} \left( \sigma_i^* \right)_{1s} a_i^\dag \alpha_s^\dag
\end{split}
\end{equation}
\end{subequations}

The remaining commutation relations of the superalgebra $\mathfrak{f}(4)$ are given below: 
\begin{subequations}
\begin{equation}
\anticommute{\mathcal{Q}_I}{\overline{\mathcal{Q}}_J}
= \Omega_{IJ} \left( 3 \, \mathcal{H} - 2 \, \mathcal{Z} \right)
  + U_{IJ}
\end{equation}
\begin{equation}
\begin{aligned}
\anticommute{\mathcal{Q}_I}{\mathcal{R}_J}
&= + 3 \, \Omega_{IJ} \, T_-
\\
\anticommute{\mathcal{Q}_I}{\overline{\mathcal{R}}_J}
&= - \mathcal{B}_{IJ}
\\
\commute{\mathcal{Z}}{\mathcal{Q}_I}
&= - 2 \, \mathcal{Q}_I
\\
\commute{U_{IJ}}{\mathcal{Q}_K}
&= \Omega_{JK} \, \mathcal{Q}_I + \Omega_{IK} \, \mathcal{Q}_J
\\
\commute{U_{IJ}}{T_+}
&= 0
\end{aligned}
\qquad \qquad
\begin{aligned}
\anticommute{\overline{\mathcal{R}}_I}{\overline{\mathcal{Q}}_J}
&= - 3 \, \Omega_{IJ} \, T_+
\\
\anticommute{\mathcal{R}_I}{\overline{\mathcal{Q}}_J}
&= + \overline{\mathcal{B}}_{IJ}
\\
\commute{\mathcal{Z}}{\overline{\mathcal{Q}}_I}
&= + 2 \, \overline{\mathcal{Q}}_I
\\
\commute{U_{IJ}}{\overline{\mathcal{Q}}_K}
&= \Omega_{JK} \, \overline{\mathcal{Q}}_I
   + \Omega_{IK} \, \overline{\mathcal{Q}}_J
\\
\commute{U_{IJ}}{T_-}
&= 0
\end{aligned}
\end{equation}
\end{subequations}

It should be noted that, among the four supersymmetry operators of $\mathfrak{C}^\pm$ subspaces, the following relations hold: 
\begin{equation}
\Omega_{IJ} \mathcal{Q}_I \mathcal{Q}_J = 0
\qquad \qquad \qquad
\Omega_{IJ} \overline{\mathcal{Q}}_I \overline{\mathcal{Q}}_J = 0
\label{SUSYconstraint}
\end{equation}


\section{Minimal unitary supermultiplet of $F(4)$}
\label{sec:F(4)supermultiplet}

In the Hilbert space spanned by the tensor product states of the form (\ref{tensorstates}), there exists a unique lowest weight vector that is annihilated by all grade $-1$ generators $\mathcal{B}_{IJ}$, $T_-$ and $\mathcal{Q}_I$ in the compact 3-grading and is a singlet of the compact subsuperalgebra $\mathfrak{osp}(2|4)$ with definite eigenvalue of $\mathcal{H}$, namely the lowest weight vector of the minrep of $SO(5,2)$ which we labelled as $|\Phi_{0,0}\rangle$ in equation (\ref{SO(5,2)LWV0_1}):
\begin{equation}
\ket{\Phi_{0,0}}
= | \psi_0^{(\alpha_g=1)} \rangle
= C_0 \, x \, e^{-x^2/2} \ket{0} \,\,,
\label{MinrepSupermultipletLWV}
\end{equation}
where $\ket{0}$ is the Fock vacuum of the $a$-type bosonic oscillators and the $\alpha$-type fermionic oscillators.

By acting on $\ket{\Phi_{0,0}}$ with the operators $\overline{\mathcal{B}}_{IJ}$, $T_+$ and $\mathcal{\overline{Q}}_I$ in grade +1 subspace $\mathfrak{C}^+$ repeatedly, one obtains an infinite set of states which forms a basis for the minimal unitary irreducible representation of $\mathfrak{f}(4)$. This infinite set of states can be decomposed into a finite number of irreducible representations of the even subgroup $SO(5,2) \times SU(2)_T$, with each irrep of $SO(5,2)$ corresponding to a massless conformal field in five dimensions.
In Table \ref{Table:minrepsupermultiplet}, we present the minimal unitary supermultiplet of $\mathfrak{f}(4)$ corresponding to this unique lowest weight vector $\ket{\Phi_{0,0}}$. We should note that the action of $\overline{\mathcal{B}}_{IJ}$ moves one within an irrep of $SO(5,2)$ and action of $T_+$ moves one within the internal symmetry  group $SU(2)_T$. On the other hand, with the action of the supersymmetry operators $\mathcal{\overline{Q}}_I$ of grade $+1$ subspace $\mathfrak{C}^+$, one moves between different irreps of $SO(5,2)$ that make up the supermultiplet.

\begin{small}
\begin{longtable}[c]{|c|c|c|c|}
\kill

\caption[Minimal unitary supermultiplet of $\mathfrak{f}(4)$]
{Minimal unitary supermultiplet of $\mathfrak{f}(4)$. The $AdS$ energy $E$, spin $t$ of the $R$-symmetry group $SU(2)_T$, and the $U\!Sp(4)$ Dynkin labels of each level are given.
\label{Table:minrepsupermultiplet}} \\
\hline
 & & & \\
States of the & $AdS$ & $SU(2)_T$ Labels  & Dynkin Labels \\
Lowest Energy & Energy & $(t,t_3)$ & of $U\!Sp(4)$ \\
$U\!Sp(4)$ Irreps & $E$ & & \\
\hline
 & & & \\
\endfirsthead
\caption[]{(continued)} \\
\hline
 & & & \\
States of the & $AdS$ & $SU(2)_T$ Labels  & Dynkin Labels \\
Lowest Energy & Energy & $(t,t_3)$ & of $U\!Sp(4)$ \\
$U\!Sp(4)$ Irreps & $E$ & & \\
\hline
 & & & \\
\endhead
 & & & \\
\hline
\endfoot
 & & & \\
\hline
\endlastfoot

$\ket{\Phi_{0,0}} $ &
$\frac{3}{2}$ &
$(\frac{1}{2},-\frac{1}{2})$
 & $(0,0)$
\\[8pt]


$\overline{\mathcal{Q}}_I \ket{\Phi_{0,0}} $ &
$2$ &
$(0,0)$
 & $(1,0)$
\\[8pt]

$T_+ \ket{\Phi_{0,0} }$ &
$\frac{3}{2}$ &
$(\frac{1}{2},+\frac{1}{2})$  & $(0,0)$

\end{longtable}
\end{small}

Therefore, interpreted as the $N=2$ superconformal algebra in five dimensions, the minimal unitary supermultiplet of $\mathfrak{f}(4)$ corresponds to a supermultiplet of fields consisting of two copies of the scalar singleton transforming as a doublet of the R-symmetry group $SU(2)_T$ and a singlet of the spinor singleton of $SO(5,2)$.\footnote{In Appendix \ref{App:Intertwiner} we give the interwiner between compact and noncompact bases and explain how to determine the massless conformal fields corresponding to the unitary irreducible representations of $SO(5,2)$.}
Labeling the doublet of massless conformal scalar fields  as $\Phi^r_{(0,0)}(x^\mu)$ ($r=1,2$) and the massless spinor field as $\Psi_{(1,0)}(x^\mu)$ we have the minimal superconformal multiplet of $F(4)$ as:
\be 
\Psi_{(1,0)}(x^\mu) \,\, \oplus \,\,  \Phi^r_{(0,0)}(x^\mu) 
\ee
where the subscripts indicate the Dynkin labels of the Lorentz group $U\!Sp(2,2)$.


\section{ $AdS_6/CFT_5$ bosonic higher spin algebra and its deformation} 
\label{sec:bosonicHS}

The $AdS_d/CFT_{(d-1)}$ higher spin algebra of Fradkin-Vasiliev type corresponds simply to the universal enveloping algebra of $SO(d,2)$ quotiented by its Joseph ideal \cite{Gunaydin:1989um,Vasiliev:1999ba,Eastwood:2002su,eastwood2005uniqueness,Govil:2013uta,Govil:2014uwa}. The Joseph ideal of a Lie algebra $\mathfrak{g}$ is a two-sided ideal that annihilates its minimal unitary representation. We shall denote the corresponding higher spin algebra as $hs(d-1,2)$:
\begin{equation}
hs(d-1,2) = \frac{\mathscr{U}(d-1,2)}{\mathscr{J}(d-1,2)} 
\end{equation}
where $\mathscr{U}(d-1,2)$ is the universal enveloping algebra of $\mathfrak{so}(d-1,2)$ and $\mathscr{J}(d-1,2)$ denotes its Joseph ideal.

An explicit formula for the generators of this ideal for $SO(d-1,2)$
was given by Eastwood \cite{eastwood2005uniqueness}:
\begin{equation}
\begin{split}
J_{ABCD}
&= M_{AB} M_{CD}
   - M_{AB} \circledcirc M_{CD}
   - \frac{1}{2} \commute{M_{AB}}{M_{CD}}
   + \frac{(d-3)}{4 d(d-1)} \langle M_{AB} , M_{CD} \rangle \, \mathbf{1}
\\
&= \frac{1}{2} M_{AB} \cdot M_{CD}
   - M_{AB} \circledcirc M_{CD}
   + \frac{(d-3)}{4d(d-1)} \langle M_{AB} , M_{CD} \rangle \,  \mathbf{1} \end{split}
\label{Joseph} 
\end{equation}
where the dot $\cdot$ denotes the symmetric product of the generators $M_{AB}$ of $\mathfrak{so}(d-1,2)$
\begin{equation}
M_{AB} \cdot M_{CD}
= M_{AB} M_{CD} + M_{CD} M_{AB} \,,
\end{equation}
$\langle M_{AB} , M_{CD} \rangle$ is the Killing form of $SO(d-1,2)$ given
by
\begin{equation}
\langle M_{AB} , M_{CD} \rangle
= \frac{2(d-1)}{(d+1)(3-d)} \, M_{EF} M_{GH}
  \left( \eta^{EG} \eta^{FH} - \eta^{EH} \eta^{FG} \right)
  \left( \eta_{AC} \eta_{BD} - \eta_{AD} \eta_{BC} \right)
\end{equation}
where $\eta^{AB}$ denotes the $SO(d-1,2)$ invariant metric, and the symbol $\circledcirc$ denotes the Cartan product of two generators \cite{eastwood2005cartan}:
\begin{equation}
\begin{split}
M_{AB} \circledcirc M_{CD}
&= \frac{1}{3} M_{AB} M_{CD}
   + \frac{1}{3} M_{DC} M_{BA}
   + \frac{1}{6} M_{AC} M_{BD}
\\
& \quad
   - \frac{1}{6} M_{AD} M_{BC}
   + \frac{1}{6} M_{DB} M_{CA}
   - \frac{1}{6} M_{CB} M_{DA}
\\
& \quad
   - \frac{1}{2(d-1)}
     \left(
      M_{AE} M_C^{~E} \eta_{BD} - M_{BE} M_C^{~E} \eta_{AD}
     \right.
\\
& \qquad \qquad \qquad
     \left.
      + M_{BE} M_D^{~E} \eta_{AC} - M_{AE} M_D^{~E} \eta_{BC}
     \right)
\\
& \quad
   - \frac{1}{2(d-1)}
     \left(
      M_{CE} M_A^{~E} \eta_{BD} - M_{CE} M_B^{~E} \eta_{AD}
     \right.
\\
& \qquad \qquad \qquad
     \left.
      + M_{DE} M_B^{~E} \eta_{AC} - M_{DE} M_A^{~E} \eta_{BC}
     \right)
\\
& \quad
   + \frac{1}{d(d-1)} M_{EF} M^{EF}
     \left( \eta_{AC} \eta_{BD} - \eta_{BC} \eta_{AD} \right)
\end{split}
\end{equation}
For $SO(5,2)$ the generator of the Joseph ideal takes the form:
\begin{equation}
J_{ABCD}
= \frac{1}{2} M_{AB} \cdot M_{CD}
  - M_{AB} \circledcirc M_{CD} 
  - \frac{1}{40} \langle M_{AB} , M_{CD} \rangle
\label{joseph-6d}
\end{equation}

As was shown by Vasiliev \cite{Vasiliev:1999ba}  the bosonic higher spin fields correspond to tensorial fields, which under the adjoint action of $SO(d-1,2)$, transform in representations whose Young tableaux have two rows only\footnote{Uniqueness of the bosonic higher spin symmetries in various dimensions using Young-tableaux techniques was studied in \cite{Boulanger:2013zza}. However their analysis does not cover the case of $AdS_6/CFT_5$ higher spin symmetries.}. On the other hand the enveloping algebra $\mathscr{U}(\mathfrak{so}(d-1,2)$ is obtained by taking symmetric tensor products of the generators that transform in the adjoint representation whose Young tableau is
\begin{equation}
M_{AB} \sim \parbox{10pt}{\YoungAA}
\end{equation}
Symmetric tensor product of the adjoint representation of $SO(5,2)$ decomposes as 
\begin{equation}
\bf (21 \times 21 )_S = 1 + 27 + 35 + 168
\end{equation}
where the singlet corresponds to the quadratic Casimir, $\bf 27$ to the symmetric traceless tensor and $\bf 35$ to the completely antisymmetric tensor of rank 4. Modding out by the Joseph ideal removes all but the representation $\bf 168$ corresponding to the window diagram
\begin{equation}
{\bf 168} \simeq \parbox{20pt}{\YoungBB}
\end{equation}
Higher symmetric products yield operators whose Young tableaux have two rows only after modding out by the Joseph ideal
\be 
\underbrace{\begin{picture}(100,50)(01.8,-18)
\put(0,5){\line(0,1){20}}
\put(00,5){\line(1,0){100}}
\put(00,15){\line(1,0){100}}
\put(00,25){\line(1,0){100}}
\put(10,5){\line(0,1){20}}
\put(20,5){\line(0,1){20}}
\put(30,5){\line(0,1){20}}
\put(40,5){\line(0,1){20}}
\put(70,5){\line(0,1){20}}
\put(80,5){\line(0,1){20}}
\put(90,5){\line(0,1){20}}
\put(100,5){\line(0,1){20}}
\put(55,10){\makebox(0,0){$\cdots$}}
\put(55,20){\makebox(0,0){$\cdots$}}
\end{picture}}_{\mbox{$n$ boxes}}
\ee

Since Eastwood's results are written in $SO(5,2)$-covariant form, let us list the generators that extend the Lie algebra of $SO(4,1)$ given in equation (\ref{lorentz}) to those of $SO(5,2)$ as given in Appendix \ref{SO(5,2)generators}:
\begin{equation}
\begin{split}
M_{05} &= \frac{1}{2} (\mathcal{P}_0-\mathcal{K}_0)
\qquad \qquad
M_{06} = \frac{1}{2} (\mathcal{P}_0+\mathcal{K}_0)
\\
M_{15} &= \frac{1}{2} (\mathcal{P}_1-\mathcal{K}_1)
\qquad \qquad
M_{16} = \frac{1}{2} (\mathcal{P}_1+\mathcal{K}_1)
\\
M_{25} &= \frac{1}{2} (\mathcal{P}_2-\mathcal{K}_2)
\qquad \qquad
M_{26} = \frac{1}{2} (\mathcal{P}_2+\mathcal{K}_2)
\\
M_{35} &= \frac{1}{2} (\mathcal{P}_3-\mathcal{K}_3)
\qquad \qquad
M_{36} = \frac{1}{2} (\mathcal{P}_3+\mathcal{K}_3)
\\
M_{45} &= \frac{1}{2} (\mathcal{P}_4-\mathcal{K}_4)
\qquad \qquad
M_{46} = \frac{1}{2} (\mathcal{P}_4+\mathcal{K}_4)
\\
M_{56} &= \mathcal{-D}
\end{split}
\end{equation}

Inserting the generators for the minimal unitary representation of $\mathfrak{so}(5,2)$ constructed using quasiconformal methods into the expression for the generator $J_{ABCD}$ of \textit{Joseph ideal one finds that it vanishes identically as an operator}. Therefore taking the enveloping algebra of the corresponding minimal unitary realization yields directly the $AdS_6/CFT_5$  higher spin algebra in complete parallel  to the situation for the $AdS_5/CFT_4$ and $AdS_7/CFT_6$ higher spin algebras \cite{Govil:2013uta,Govil:2014uwa}. The generators of the infinite higher spin algebra then decompose under the finite dimensional subalgebra $\mathfrak{so}(5,2)$ as 
\be
\parbox{10pt}{\YoungAA} \quad \oplus \quad  \parbox{20pt}{\YoungBB} \quad \oplus \quad ...\quad \oplus\quad 
\begin{picture}(100,50)(01.8,13)
\put(0,5){\line(0,1){20}}
\put(00,5){\line(1,0){100}}
\put(00,15){\line(1,0){100}}
\put(00,25){\line(1,0){100}}
\put(10,5){\line(0,1){20}}
\put(20,5){\line(0,1){20}}
\put(30,5){\line(0,1){20}}
\put(40,5){\line(0,1){20}}
\put(70,5){\line(0,1){20}}
\put(80,5){\line(0,1){20}}
\put(90,5){\line(0,1){20}}
\put(100,5){\line(0,1){20}}
\put(55,10){\makebox(0,0){$\cdots$}}
\put(55,20){\makebox(0,0){$\cdots$}}
\end{picture} \quad \oplus\quad ....
\ee

For the deformed minrep of $SO(5,2)$, obtained above via quasiconformal methods, the Joseph ideal does not vanish identically as an operator. More specifically for the deformed minimal unitary representation, the symmetric traceless operator occurring in the tensor product of two generators of $\mathfrak{so}(5,2)$ corresponding to the tableau $\parbox{20pt}{\YoungB}$ still vanishes:
\begin{equation}
\eta^{CD} M_{AC} \cdot M_{DB} = 0
\end{equation}
However the operator corresponding to the Young tableaux $\parbox{10pt}{\YoungAAAA}$ does not vanish for the deformed minimal unitary representation.
To see how the Joseph ideal gets deformed, we rewrite the generators of
the Joseph ideal in a $5d$ Lorentz-covariant form as was done for the $AdS_5/CFT_4$ and $AdS_7/CFT_6$ higher spin algebras in \cite{Govil:2013uta,Govil:2014uwa}.

First we have the conditions:
\begin{equation}
P^2 = \mathcal{P}^\mu \mathcal{P}_\mu = 0
\quad , \quad
K^2 = \mathcal{K}^\mu \mathcal{K}_\mu = 0
\end{equation}
which are valid for both the scalar minrep as well as the spinorial minrep of
$SO(5,2)$. The quadratic relations that define the Joseph ideal in the $SO(4,1)$-covariant basis are as follows:
\begin{eqnarray}
5 \, \mathcal{D} \cdot \mathcal{D}
+ \mathcal{M}^{\mu\nu} \cdot \mathcal{M}_{\mu\nu}
+ \frac{3}{2} \mathcal{P}^\mu \cdot \mathcal{K}_\mu
&=& 0
\\
\mathcal{P}^\mu \cdot
\left( \mathcal{M}_{\mu\nu} + \eta_{\mu\nu} \,\mathcal{D} \right)
&=& 0
\\
\mathcal{K}^\mu \cdot
\left( \mathcal{M}_{\nu\mu} - \eta_{\nu\mu} \, \mathcal{D} \right)
&=& 0
\\
\eta^{\mu\nu} \mathcal{M}_{\mu\rho} \cdot \mathcal{M}_{\nu\sigma}
- \mathcal{P}_{(\rho} \cdot \mathcal{K}_{\sigma)}
+ 3 \, \eta_{\rho\sigma}
&=& 0 \label{mpident}
\\
\mathcal{M}_{\mu\nu} \cdot \mathcal{M}_{\rho\sigma}
+ \mathcal{M}_{\mu\sigma} \cdot \mathcal{M}_{\nu\rho}
+ \mathcal{M}_{\mu\rho} \cdot \mathcal{M}_{\sigma\nu}
&=& 0  \label{mident}\\
\mathcal{D} \cdot \mathcal{M}_{\mu\nu}
+ \mathcal{P}_{[\mu} \cdot \mathcal{K}_{\nu]}
&=& 0
\\
\mathcal{M}_{[\mu\nu} \cdot \mathcal{P}_{\rho]}
&=& 0
\\
\mathcal{M}_{[\mu\nu} \cdot \mathcal{K}_{\rho]}
&=& 0 
\end{eqnarray}
where the dilatation  generator $\mathcal{D}$ is identified as $M_{65}$ and symmetrizations (round brackets) and anti-symmetrizations (square brackets) are done with weight one.\footnote{We should stress that the dot product of two operators $A$ and $B$ is defined as $A \cdot B = AB + BA$.} The above identities are all valid for the minimal unitary realization of $SO(5,2)$. However when we substitute the expressions for the generators of the deformed minrep of $SO(5,2)$ involving fermionic oscillators, one finds that only the first three equations above remain unchanged. The fourth equation above gets modified as follows:
\begin{equation}
\eta^{\mu\nu} \mathcal{M}_{\mu\rho} \cdot \mathcal{M}_{\nu\sigma}
- \mathcal{P}_{(\rho} \cdot \mathcal{K}_{\sigma)}
- \left( \frac{4}{3} \mathcal{S}^2 - 3 \right) \eta_{\rho\sigma}
= 0 
\end{equation}
where $\mathcal{S}^2$ is the Casimir of $SU(2)_S$ realized as bilinears of the fermionic oscillators. Similarly, the remaining four identities above get modified by spin-dependent terms involving fermionic oscillators. 

By replacing the fermionic oscillator realization of $SU(2)_S$ generators by the two dimensional Pauli matrices $S_i = \sigma_i/2$ above one obtains the spinor singleton realization by itself whose enveloping algebra then defines a deformed $AdS_6/CFT_5$ higher spin algebra for which the deformed Joseph ideal vanishes identically.


\section{ Unique $AdS_6/CFT_5$ higher spin superalgebra as enveloping algebra of the minimal unitary representation of $F(4)$}
\label{sec:higherspin}

Any definition of a higher spin superalgebra must have, as a subalgebra, the bosonic higher spin algebra based on the even subalgebra of the underlying finite-dimensional subsuperalgebra. Hence shall define the higher spin $AdS_6/CFT_5$ superalgebra as the enveloping algebra of the quasiconformal realization of the minimal unitary representation of the superalgebra $\mathfrak{f}(4)$, which  has as subalgebras the bosonic  higher spin algebra defined by the scalar singleton as well as the deformed higher spin algebra defined by the spinor singleton in five dimensions. To exhibit  the structure of the resulting higher spin superalgebra we shall reformulate the minimal unitary realization of $\mathfrak{f}(4)$ in an $SO(5,2)$-covariant basis.

For this  we first choose  the $SO(5,2)$ gamma-matrices $\left( \Gamma_A \right)^\alpha_{~\beta}$ ($A = 0,\dots,6$) as follows:
\begin{equation}
\begin{split}
\left( \Gamma_0 \right)^\alpha_{~\beta}
&= i \, \sigma_2 \otimes \mathbb{I}_4
 = \left(
    \begin{matrix} 0 & \mathbb{I}_4 \\ - \mathbb{I}_4 & 0 \end{matrix}
   \right)
\\
\left( \Gamma_M \right)^\alpha_{~\beta}
&= \sigma_3 \otimes \gamma_M
 = \left(
    \begin{matrix} \gamma_M & 0 \\ 0 & - \gamma_M \end{matrix}
   \right)
\\
\left( \Gamma_6 \right)^\alpha_{~\beta}
&= - i \, \sigma_1 \otimes \mathbb{I}_4
 = \left(
    \begin{matrix} 0 & - i \, \mathbb{I}_4 \\ - i \, \mathbb{I}_4 & 0 \end{matrix}
   \right)
\end{split}
\end{equation}
where $\gamma_M$ are the $SO(5)$ gamma-matrices given in equation (\ref{SO(5)gamma}), $M=1,\dots,5$, and $\alpha,\beta=1,\dots,8$ are the spinor indices of $SO(5,2)$.  The symmetric $SO(5,2)$ charge conjugation matrix is chosen to be
\begin{equation}
\left( \mathcal{C}_7 \right)_{\alpha\beta}
= \left( \mathcal{C}_7 \right)^{\alpha\beta}
= - \sigma_2 \otimes \mathbb{I}_2 \otimes \sigma_2
= i \, \sigma_2 \otimes \mathcal{C}_5
\end{equation}
With the above choices, we find that all the matrices $\left( \mathcal{C}_7 \Gamma_A \right)_{\alpha\beta}$ and $\left( \mathcal{C}_7 \commute{\Gamma_A}{\Gamma_B} \right)_{\alpha\beta}$ are antisymmetric. The spinor representation of $SO(5,2)$ is then realized by the matrices
\begin{equation}
\Sigma_{AB} = \frac{i}{4} \commute{\Gamma_A}{\Gamma_B} \,.
\end{equation}
which satisfy  commotion relations
\begin{equation}
\commute{\Sigma_{AB}}{\Sigma_{CD}}
= i \left(
     \eta_{BC} \Sigma_{AD} - \eta_{AC} \Sigma_{BD}
     - \eta_{BD} \Sigma_{AC} + \eta_{AD} \Sigma_{BC}
    \right)
\end{equation}
where $\eta_{AB} = \mathrm{diag} \left( -,+,+,+,+,+,- \right)$.

The 16 supersymmetry generators of $F(4)$ transform as two eight dimensional spinors $\Xi_\alpha$, $\overline{\Xi}_\alpha$ of $SO(5,2)$ and are defined as 
\begin{equation}
\Xi_\alpha
= \left(
   \begin{matrix}
    \mathcal{Q}_I \\ - i \, \mathcal{R}_I
   \end{matrix}
  \right)
\qquad \qquad \qquad
\overline{\Xi}_\alpha
= \left(
   \begin{matrix}
    \overline{\mathcal{R}}_I \\ - i \, \overline{\mathcal{Q}}_I
   \end{matrix}
  \right)
\end{equation}
where $I=1,2,3,4$ is the  $U\!Sp(4)$ spinor index.
Under commutation with  $SO(5,2)$ generators $M_{AB}$ they satisfy:
\begin{equation}
\commute{M_{AB}}{\Xi_\alpha}
= - \left( \Sigma_{AB} \right)_{\alpha\beta} \Xi_\beta
\qquad \qquad
\commute{M_{AB}}{\overline{\Xi}_\alpha}
= - \left( \Sigma_{AB} \right)_{\alpha\beta} \overline{\Xi}_\beta
\end{equation}

Under anticommutation, they close into $SO(5,2)$ and $SU(2)_T$ generators as follows:
\begin{equation}
\begin{split}
\anticommute{\Xi_\alpha}{\Xi_\beta}
&= - 3 i \, \left( \mathcal{C}_7 \right)_{\alpha\beta} T_-
\\
\anticommute{\overline{\Xi}_\alpha}{\overline{\Xi}_\beta}
&= + 3 i \, \left( \mathcal{C}_7 \right)_{\alpha\beta} T_+
\\
\anticommute{\Xi_\alpha}{\overline{\Xi}_\beta}
&= i \, M_{AB} \left( \Sigma^{AB} \mathcal{C}_7 \right)_{\alpha\beta}
   + 3 i \, \left( \mathcal{C}_7 \right)_{\alpha\beta} T_3
\end{split}
\end{equation}

 Defining the $SU(2)_T$ doublet of $SO(5,2)$ supersymmetry generators 
\begin{equation}
\overline{\Xi}_\alpha = \Xi_\alpha^1
\qquad
\Xi_\alpha = \Xi_\alpha^2
\end{equation}
the above anticommutation relations can be recast in an $SU(2)_T$ covariant form:
\begin{equation}
\anticommute{\Xi_\alpha^r}{\Xi_\beta^s}
=  i \epsilon^{rs} \, M_{AB} \left( \Sigma^{AB} \mathcal{C}_7 \right)_{\alpha\beta}
   + 3 i \, \left( \mathcal{C}_7 \right)_{\alpha\beta} ( i\sigma_2 \sigma^i)^{rs} \, T_i
\end{equation}
where $\epsilon^{rs}$ is the two dimensional Levi-Civita tensor and $r,s=1,2$ are the $SU(2)_T$ spinor indices. The expressions for the $SO(5,2)$ generators $M_{AB}$ in terms of the generators in the noncompact 3-grading and the compact 3-grading are collected in Appendix \ref{App:SO(5,2)}.

The higher spin $AdS_6/CFT_5$ algebra defined by $\mathfrak{f}(4)$ has, as a subalgebra, the enveloping algebra of the $SO(5,2) $ subalgebra spanned by symmetric products of the  generators $M_{AB}$. It also has, as a subalgebra, the enveloping algebra of $SU(2)_T$ spanned by the symmetric products of the generators $T_i$, which is finite-dimensional and consists of $T_i$ and the quadratic Casimir element $\mathcal{T}^2$. The additional even elements of the higher spin superalgebra are given by products of the elements of the enveloping algebras of $SO(5,2)$ and of $SU(2)_T$  with antisymmetric products of an even number of supersymmetry generators $\Xi^r_\alpha$. The odd elements of the higher spin algebra are given by products of the generators of the enveloping algebras of $SO(5,2)$ and $SU(2)_T$ with antisymmetric products of an odd number of supersymmetry generators $\Xi^r_\alpha$.


\section{ Comments} 
\label{sec:comments}

The results obtained in this paper on the minimal unitary representations of $SO(5,2)$ and  of the exceptional superalgebra $F(4)$  and the $AdS_6/CFT_5$ higher spin (super)algebras can be  further developed and applied in several directions.   First is the construction of  bosonic higher theory in $AdS_6$ in terms of covariant fields based on the minrep of $SO(5,2)$ and its supersymmetric extension  based on the minrep of $F(4)$. Our results are also relevant  to $AdS_6/CFT_5$ dualities studied in references \cite{Ferrara:1998gv,D'Auria:2000ad,D'Auria:2000ah,D'Auria:2000ay} and their extensions to  higher spin theories. Yet another application is to integrable spin chain models related to conformally invariant (super)-symmetric field theories in five dimensions. \\

{\bf Acknowledgements:} 
One of us (M.G.) would like to thank the CERN Theory Division and the Albert Einstein Institute for their hospitality where part of this work was done. The research of M.G. is supported in part by the US Department of Energy under DOE Grant No: DE-SC0010534.
S.F. would like to thank the Center for Fundamental Theory of  the Institute for Gravitation and the Cosmos at Pennsylvania State University, where part of this work was done.
We would like to thank Karan Govil, Evgeny Skvortsov and Massimo Taronna for stimulating discussions regarding higher spin algebras.

\newpage

\appendix

\numberwithin{equation}{section}

\section*{Appendices}


\section{Relations between the generators in noncompact 3-grading and compact 3-grading}
\label{App:SO(5,2)}

The generators $M_{AB}$ ($A,B = 0,\dots,6$) of $\mathfrak{so}(5,2)$ algebra can be expressed in terms of the generators in the noncompact 3-grading (\emph{i.e.} five-dimensional conformal generators) $\mathcal{D}$, $\mathcal{M}_{\mu\nu}$, $\mathcal{P}_\mu$, $\mathcal{K}_\mu$ ($\mu,\nu = 0,\dots,4$) and the generators in the compact 3-grading $H$, $\widetilde{M}_{MN}$, $\widetilde{B}^\dag_M$, $\widetilde{B}_M$ ($M,N = 1,\dots,5$) as follows:
\begin{equation}
\begin{split}
M_{0i}
&= \mathcal{M}_{0i}
 = \frac{1}{2} \left( \widetilde{B}_i^\dag + \widetilde{B}_i \right)
\\
M_{04}
&= \mathcal{M}_{04}
 = \frac{1}{2} \left( \widetilde{B}_4^\dag + \widetilde{B}_4 \right)
\\
M_{ij}
&= \mathcal{M}_{ij}
 = \widetilde{M}_{ij}
\\
M_{i4}
&= \mathcal{M}_{i4}
 = \widetilde{M}_{i4}
\\
M_{05}
&= \frac{1}{2} \left( \mathcal{P}_0 - \mathcal{K}_0 \right)
 = \frac{1}{2} \left( \widetilde{B}_5^\dag + \widetilde{B}_5 \right)
\\
M_{i5}
&= \frac{1}{2} \left( \mathcal{P}_i - \mathcal{K}_i \right)
 = \widetilde{M}_{i5}
\\
M_{45}
&= \frac{1}{2} \left( \mathcal{P}_4 - \mathcal{K}_4 \right)
 = \widetilde{M}_{45}
\\
M_{56}
&= - \mathcal{D}
 = \frac{i}{2} \left( \widetilde{B}_5^\dag - \widetilde{B}_5 \right)
\\
M_{06}
&= \frac{1}{2} \left( \mathcal{P}_0 + \mathcal{K}_0 \right)
 = H
\\
M_{i6}
&= \frac{1}{2} \left( \mathcal{P}_i + \mathcal{K}_i \right)
 = \frac{i}{2} \left( \widetilde{B}_i^\dag - \widetilde{B}_i \right)
\\
M_{46}
&= \frac{1}{2} \left( \mathcal{P}_4 + \mathcal{K}_4 \right)
 = \frac{i}{2} \left( \widetilde{B}_4^\dag - \widetilde{B}_4 \right)
\end{split}
\label{SO(5,2)generators}
\end{equation}
Note that $i,j=1,2,3$.

They satisfy the canonical commutation relations
\begin{equation}
\commute{M_{AB}}{M_{CD}}
= i \left(
    \eta_{BC} M_{AD} - \eta_{AC} M_{BD} - \eta_{BD} M_{AC} + \eta_{AD} M_{BC}
    \right)
\end{equation}
where $\eta_{AB} = \mathrm{diag} \left( -,+,+,+,+,+,- \right)$.


\section{The ``intertwiner'' between compact and non-compact  3-graded bases of $SO(5,2)$ }
\label{App:Intertwiner}

 
Consider the ``intertwiner'' operator
\begin{equation}
T = e^{\frac{\pi}{2} M_{05}}
\end{equation}
where $M_{05}$, as defined in equation (\ref{SO(5,2)generators}), is simply
$M_{05} = \frac{1}{2} \left( \mathcal{P}_0 - \mathcal{K}_0 \right)$.
When acting on $SO(5,2)$ generators in the manifestly unitary  compact three-grading
\begin{equation}
\mathfrak{so}(5,2)_c
= \widetilde{B}_M \,\, \oplus \,\,
  \left[ H \,,\, \widetilde{M}_{MN} \right] \,\, \oplus \,\,
  \widetilde{B}_M^\dag
\end{equation}
this operator  $T$ intertwines them into the corresponding generators in the five-dimensional, manifestly Lorentz-covariant, noncompact three-graded basis 
\begin{equation}
\mathfrak{so}(5,2)_{nc}
= \mathcal{K}_\mu \,\, \oplus \,\,
  \left[ \mathcal{D} \,,\, \mathcal{M}_{\mu\nu} \right] \,\, \oplus \,\,
  \mathcal{P}_\mu
\end{equation}
More specifically we have 
\begin{equation}
\begin{aligned}
T \, \widetilde{B}_M \, T^{-1}
&\longrightarrow
\mathcal{K}_\mu
\\
T \, \widetilde{B}_M^\dag \, T^{-1}
&\longrightarrow
\mathcal{P}_\mu
\end{aligned}
\qquad \qquad \qquad
\begin{aligned}
T \, \widetilde{M}_{MN} \, T^{-1}
&\longrightarrow
\mathcal{M}_{\mu\nu}
\\
T \, H \, T^{-1}
&\longrightarrow
\mathcal{D}
\end{aligned}
\end{equation}
such that $M,N=1,2,3,4 \, \longrightarrow \, \mu,\nu=1,2,3,4$ and $M=5 \, \longrightarrow \, \mu=0$.

Acting with $T$ on states $| \Omega_{(m,n)}(0) \rangle$ that transform covariantly under $U\!Sp(4)$ with Dynkin labels $(m,n)$ in the compact 3-grading, one obtains states $| \Phi_{(m,n)}(0) \rangle$ that transform covariantly under the Lorentz group $USp(2,2)$ with the same Dynkin labels
\begin{equation}
T \ket{\Omega_{(m,n)}(0)} \, \longrightarrow \, \ket{\Phi_{(m,n)}(0)}
\end{equation}
For states $\ket{\Omega_{(m,n)}(0)}$ that belong to the lowest energy irrep, one finds  that the corresponding state $\ket{\Phi_{(m,n)}(0)}=T\ket{\Omega_{(m,n)}(0)}$ in the noncompact picture is annihilated by the special conformal generators: 
\begin{equation}
\widetilde{B}_M \ket{\Omega_{(m,n)}(0)} = 0
\quad \Longrightarrow \quad
T \, \widetilde{B}_M \, T^{-1} \, T \ket{\Omega_{(m,n)}(0)} = 0
\, \longrightarrow \,
\mathcal{K}_\mu \ket{\Phi_{(m,n)}(0)} = 0
\end{equation}
Then by acting on these covariant states $\ket{\Phi_{(m,n)}(0)}$ with $e^{-i x^\mu P_\mu}$, one forms coherent states that are labelled by the coordinates $x^\mu$:
\begin{equation}
e^{-i x^\mu P_\mu} \ket{\Phi_{(m,n)}(0)} = \ket{\Phi_{(m,n)} \left( x^\mu \right)}
\end{equation}
These coherent states transform exactly like the states generated by the action of conformal fields $\Phi_{(m,n)} \left( x^\mu \right)$ on the vacuum $\ket{0}$:
\begin{equation}
\Phi_{(m,n)} \left( x^\mu \right) \ket{0} = \ket{\Phi_{(m,n)} \left( x^\mu \right)}
\end{equation}


\section{Compact 5-grading of $\mathfrak{f}(4)$ with respect to the subalgebra $\mathfrak{usp}(4) \oplus \mathfrak{su}(2) \oplus \mathfrak{u}(1)$}
\label{App:F(4)_C5G}

With respect to the compact generator
\begin{equation}
H = \frac{1}{2} \left( 3 \, \mathcal{H} - \mathcal{Z} \right)
\end{equation}
where $H$ is the conformal Hamiltonian, $\mathfrak{f}(4)$ has a 5-grading as follows:
\begin{equation}
\begin{split}
\mathfrak{f}(4)
&= \mathfrak{C}^{(-1)} \oplus
   \mathfrak{C}^{(-1/2)} \oplus
   \left[
    \mathfrak{usp}(4) \oplus \mathfrak{su}(2)_T \oplus  \mathfrak{u}(1)_H
   \right] \oplus
   \mathfrak{C}^{(+1/2)} \oplus
   \mathfrak{C}^{(+1)}
\\
&= \mathcal{B}_{IJ} \oplus
   \left( \, \mathcal{Q}_I \,,\, \overline{\mathcal{R}}_I \, \right) \oplus
   \left[ \, H \,,\, U_{IJ} \,,\, T_{\pm,0} \, \right] \oplus
   \left( \, \overline{\mathcal{Q}}_I \,,\, \mathcal{R}_I \, \right) \oplus
   \overline{\mathcal{B}}_{IJ}
\end{split}
\end{equation}


\section{Noncompact 5-grading of $\mathfrak{f}(4)$ with respect to the subalgebra $\mathfrak{so}(4,1) \oplus \mathfrak{su}(2)_T \oplus \mathfrak{so}(1,1)$}
\label{App:F(4)_NC5G}

With respect to the dilatation generator
\begin{equation*}
\mathcal{D} = \frac{1}{2} \left[ \Delta - i \left( M_+ - M_- \right) \right]
\end{equation*}
given in equation (\ref{Dilatation}), the five-dimensional conformal superalgebra $\mathfrak{f}(4)$ has  a noncompact 5-grading in a manifestly $SO(4,1)$ covariant form:
\begin{equation}
\begin{split}
\mathfrak{f}(4)
&= \mathfrak{N}^{(-1)} \oplus
   \mathfrak{N}^{(-1/2)} \oplus
   \left[
    \mathfrak{so}(4,1) \oplus \mathfrak{su}(2)_T \oplus \mathfrak{so}(1,1)_\mathcal{D} 
   \right] \oplus
   \mathfrak{N}^{(+1/2)} \oplus
   \mathfrak{N}^{(+1)}
\\
&= \mathcal{K}_\mu \,\, \oplus \,\,
   \mathfrak{R}_{Ir} \,\,  \oplus \,\,
   \left[ \, \mathcal{D} \,,\, \mathcal{M}_{\mu\nu} \,,\, T_{\pm,0} \, \right]
    \,\, \oplus \,\,
   \mathfrak{Q}_{Ir} \,\, \oplus \,\,
   \mathcal{P}_\mu
\end{split}
\end{equation}
where $\mu,\nu = 0,1,2,3,4$; $I = 1,2,3,4$; and $r = 1,2$. In this basis, $\mathcal{M}_{\mu\nu}$ are the generators of the five-dimensional Lorentz group $Spin(4,1) \approx U\!Sp(2,2)$. Grade $-1$ generators $\mathcal{K}_\mu$ are the special conformal generators   and grade $+1$ generators $\mathcal{P}_\mu$ are the translations in five dimensions. The generators $\mathfrak{R}_{Ir}$ of grade $-\frac{1}{2}$ subspace are the special conformal supersymmetry generators, and the generators $\mathfrak{Q}_{Ir}$ of grade $+\frac{1}{2}$ subspace are the Poincar\'{e} supersymmetry generators.


\section{Indices used in the paper}
\label{App:Indices}

Here we give a list of indices we used in this paper and their ranges:
\begin{equation}
\begin{aligned}
i,j,k,l &= 1,2,3
\\
a,b,c,d &= 1,2
\\
\mu,\nu,\rho,\tau &= 0,1,2,3,4
\\
M,N,P,Q &= 1,2,3,4,5
\\
I,J,K,L &= 1,2,3,4
\\
A,B,C,D,E,F &= 0,\dots,6
\\
r,s &= 1,2
\\
\alpha,\beta &= 1,\dots,8
\end{aligned}
\qquad 
\begin{aligned}
&\mbox{$SU(2)_L$ vector indices}
\\
&\mbox{$Sp(2,\mathbb{R})$ spinor indices}
\\
&\mbox{$SO(4,1)$ vector indices}
\\
&\mbox{$SO(5)$ vector indices}
\\
&\mbox{$U\!Sp(4)$ or $U\!Sp(2,2)$  spinor indices}
\\
&\mbox{$SO(5,2)$ vector indices}
\\
&\mbox{$SU(2)_T$ spinor indices}
\\
&\mbox{$SO(5,2)$ spinor indices}
\end{aligned}
\end{equation}
\providecommand{\href}[2]{#2}\begingroup\raggedright\endgroup


\end{document}